\documentclass[usenatbib]{mn2e}
\DeclareFontFamily{U}{euc}{}
\DeclareFontShape{U}{euc}{m}{n}{<-6>eurm5<6-8>eurm7<8->eurm10}{}%
\DeclareSymbolFont{AMSc}{U}{euc}{m}{n} 
\DeclareMathSymbol{\upmu}{\mathord}{AMSc}{"16}

\usepackage{amsfonts,amssymb,amsbsy,amsmath}
\usepackage{bigints}
\usepackage{booktabs}
\usepackage{color}
\usepackage{footnote}
\usepackage{graphics}                    
\usepackage{graphicx}
\usepackage{hyperref} 
\usepackage{mathrsfs}
\usepackage{mathtools}
\usepackage{natbib}
\usepackage{nicefrac}
\usepackage{rotating}
\usepackage{slashbox}
\usepackage{subfig}
\usepackage[theorems]{tcolorbox}
\usepackage{topcapt}
\usepackage{upgreek}
\usepackage{url}
\usepackage{xifthen}

\catcode`_=\active
\newcommand_[1]{\ensuremath{\sb{\mathrm{#1}}}}

\newcommand{\newpar}{{\\}}

\newcommand{\bs}{\boldsymbol}

\newcommand{\numChar}{{n}} 

\newcommand{\paraChar}{{p}}

\newcommand{\episChar}{{n}}

\newcommand{\npe}[1][]{{ \numChar_{ \ifthenelse{\isempty{#1}}{\paraChar\episChar}{{\paraChar\episChar,#1}} } }} 

\newcommand{\epis}{{nois}}

\def\gtrsim{\mathrel{\hbox{\rlap{\hbox{\lower4pt\hbox{$\sim$}}}\hbox{$>$}}}}
\def\lessim{\mathrel{\hbox{\rlap{\hbox{\lower4pt\hbox{$\sim$}}}\hbox{$<$}}}}

\newcommand{\rmz}{{\rm z}}

\newcommand{\emodel}[1][]{{ M_{ \ifthenelse{\isempty{#1}}{\epis}{{\epis,#1}} } }}
\newcommand{\emodelz}[1][]{{ M_{ \rmz\ifthenelse{\isempty{#1}}{\epis}{{\epis,#1}} } }}
\newcommand{\emodellgrb}[1][]{{ M^\lgrb_{ \ifthenelse{\isempty{#1}}{\epis}{{\epis,#1}} } }}

\newcommand{\lgrb}{{\rm g}}

\newcommand{\param}{{\bs{\theta}}}

\newcommand{\eparam}[1][]{{ \param_{ \ifthenelse{\isempty{#1}}{\epis}{{\epis,#1}} } }}
\newcommand{\eparamz}[1][]{{ \bs\param^\rmz_{ \ifthenelse{\isempty{#1}}{\epis^\rmz}{{\epis,#1}} } }}
\newcommand{\eparamlgrb}[1][]{{ \bs\param^\lgrb_{ \ifthenelse{\isempty{#1}}{\epis^\lgrb}{{\epis,#1}} } }}

\newcommand{\truth}{{\bs{R}}}
\newcommand{\possible}{{*}}
\newcommand{\truthset}{{\mathcal{R}}}

\newcommand{\truthsubset}[1][]{{ \truthset_{ \ifthenelse{\isempty{#1}}{\truth}{{\truth_{#1}}} } }}
\newcommand{\ptruthsubset}[1][]{{ \truthset_{ \ifthenelse{\isempty{#1}}{\truth}{{\truth_{#1}}} }^\possible }}

\newcommand{\xx}[1][]{{ \ifthenelse{\isempty{#1}}{\textcolor{red}{XXX}}{\textcolor{red}{~(XXX {#1} XXX)~}} }}

\newcommand{\liso}{{L_{iso}}}
\newcommand{\eiso}{{E_{iso}}}
\newcommand{\epkz}{{E_{pz}}}
\newcommand{\durz}{{T_{90z}}}

\newcommand{\epk}{{E_{p}}}

\newcommand{\mz}{{\dot\zeta}}

\defcitealias{hopkins2006normalization}{H06}
\defcitealias{li2008star}{L08}
\defcitealias{butler2010cosmic}{B10}
\defcitealias{madau2017radiation}{M17}
\defcitealias{fermi2018gamma}{F18}

\defcitealias{yu2015unexpectedly}{Y15}
\defcitealias{petrosian2015cosmological}{P15}
\defcitealias{pescalli2016rate}{P16}
\defcitealias{tsvetkova2017konus}{T17}
\defcitealias{lloyd2019cosmological}{L19}

\title[LGRB energetics-redshift evolution]{How unbiased statistical methods lead to biased scientific discoveries: A case study of the Efron-Petrosian statistic applied to the luminosity-redshift evolution of Gamma-Ray Bursts}
\author[C. Bryant, J. A. Osborne and A. Shahmoradi]{
    Christopher Bryant$^{1}$\thanks{E-mail: christopher.bryant@mavs.uta.edu  (CMB)}
    Joshua Alexander Osborne $^{1}$\thanks{E-mail: joshua.osborne@uta.edu (JAO)}
    Amir Shahmoradi$^{1,2}$\thanks{E-mail: a.shahmoradi@uta.edu (AS) (corresponding author)}
    \\
    $^{1}$Department of Physics, College of Science, The University of Texas, Arlington, TX 76010, USA \\
    $^{2}$Data Science Program, College of Science, The University of Texas, Arlington, TX 76010, USA \\
}

\begin{document}

\date{\date{Accepted ... Received ... ; in original form ...}}

\pagerange{\pageref{firstpage}--\pageref{lastpage}} \pubyear{2020}

\maketitle

\label{firstpage}

\begin{abstract}
    Statistical methods are frequently built upon assumptions that limit their applicability to certain problems and conditions. Failure to recognize these limitations can lead to conclusions that may be inaccurate or biased. An example of such methods is the non-parametric Efron-Petrosian test statistic used in the studies of truncated data. We argue and show how the inappropriate use of this statistical method can lead to biased conclusions when the assumptions under which the method is valid do not hold. We do so by reinvestigating the evidence recently provided by multiple independent reports on the evolution of the luminosity/energetics distribution of cosmological Long-duration Gamma-Ray Bursts (LGRBs) with redshift. We show that the effects of detection threshold has been likely significantly underestimated in the majority of previous studies. This underestimation of detection threshold leads to severely-incomplete LGRB samples that exhibit strong apparent luminosity-redshift or energetics-redshift correlations. We further confirm our findings by performing extensive Monte Carlo simulations of the cosmic rates and the luminosity/energy distributions of LGRBs and their detection process.
\end{abstract}


\section{Introduction}
\label{sec:intro}

    Gamma-Ray Bursts (GRBs) are the most violent and energetic stellar explosions in the known universe. They radiate huge amounts of gamma-ray energy, comparable to the lifetime energy output of the sun, over a short period of time and are often followed by an afterglow at longer wavelengths \citep[e.g.,][]{meszaros2006gamma, zhang2007gamma, gehrels2009gamma}. With their energy concentrated in a collimated beam, they can be seen at much higher redshifts than supernovae (SNe). Amongst other things, GRBs are excellent tools to probe the Star Formation Rate (SFR) of the early as well as the recent universe.
    \newpar

    GRBs are generally divided into two categories: Long-soft GRBs (LGRBs) with $T_{90} \gtrsim 2~[s]$, and Short-hard GRBs (SGRBs) with $T_{90} \lesssim 2~[s]$.\footnote{$T_{90}$ is the duration over which a burst emits from 5-95\% of its total measured gamma-ray photon flux.} These values are based on population statistics of the \emph{Compton Gamma Ray Observatory}'s Burst and Transient Source Experiment (BATSE) detector, which was decommissioned in 2000 \citep[][]{ kouveliotou1993identification, shahmoradi2011possible, shahmoradi2013multivariate, shahmoradi2013gamma, shahmoradi2015short}. LGRBs are believed to be the result of the collapse of massive stars into a black hole \citep{woosley1993gamma}, while SGRBs are theorized to be the result of the merger of two neutron stars or of a neutron star and a black hole \citep{eichler1989nucleosynthesis}.
    \newpar

    Current research attempts to infer an accurate description and distribution profile of various GRB characteristics, in particular, the class of LGRBs due to their abundant redshift measurements compared to the SGRB class. A recent focus in the community has been on the potential cosmological evolution of LGRB luminosity/energetics $\liso/\eiso$ with redshift, as well as estimating the cosmic rates of LGRBs. A popular technique to constrain these is based on the non-parametric method of Efron-Petrosian \citep{efron1992simple, petrosian2002new}, which is widely used to study observational data sets subject to truncation and censorship \citep[e.g.,][]{kocevski2006quantifying, singal2011radio, dainotti2013determination}.
    \newpar

    \citet{yu2015unexpectedly} (hereafter \citetalias{yu2015unexpectedly}), \citet{petrosian2015cosmological} (hereafter \citetalias{petrosian2015cosmological}), and \citet{lloyd2019cosmological} (hereafter \citetalias{lloyd2019cosmological}) use the method of Efron-Petrosian to deduce the local (redshift-decorrelated) luminosity function $\psi(L_{0})$ and cosmic GRB formation rate $\rho(z)$ to infer that local GRBs $(z<1)$ are in excess of the SFR. \citet{pescalli2016rate} (hereafter \citetalias{pescalli2016rate}) follows the same approach as the previous three, however, does not find an excess of GRBs relative to the SFR at low redshifts. \citet{tsvetkova2017konus} (hereafter \citetalias{tsvetkova2017konus}) simply deduces the luminosity function and GRB formation rate.
    \newpar

    In this work, we hypothesize and provide evidence that the effects of detection threshold in the aforementioned studies might have been significantly underestimated. This underestimation of the detector threshold results in an artificial correlation between the luminosity/energetics of LGRBs and redshifts. This could, in turn, lead to the conclusion that the GRB formation rate is different from the SFR at any redshift.
    \newpar

    This paper is organized as follows. Section \ref{sec:methods} details the methodology used by the aforementioned papers to deduce luminosity/energy evolution $L(z)/E(z)$ given a sample of GRBs, which leads to $\psi(L_{0})$ and $\rho(z)$. Section \ref{sec:analysis} is a reanalysis of their work (with the exception of \citetalias{petrosian2015cosmological}, for whom we could not locate their data). Section \ref{sec:simulation} describes our own Monte Carlo simulation of a synthetic population of LGRBs. Finally, section \ref{sec:discussion} is a discussion of the results of our reanalysis and the implications of our Monte Carlo simulations. 
\section{Methods}
\label{sec:methods}

    It is often the case in the analysis of astronomical data that one is faced with reconstructing the joint bivariate or multivariate distributions from truncated data. Truncation can be due to a multitude of factors, most importantly, the Malmquist-types of biases in the population studies of GRBs \citep[e.g.,]{shahmoradi2009real, shahmoradi2011possible, shahmoradi2013multivariate, shahmoradi2015short}. The GRB luminosity and redshift distribution $(L,z)$ is one such set of bivariate data. For simplicity, it is often assumed that that the functional form of the total Luminosity Function is separable in the following form,

    \begin{equation}
        \label{eq:psi}
        \Psi(z,L) = \rho(z)\psi(L)
    \end{equation}

    \noindent where $\rho(z)$ is the GRB Formation Rate and $\psi(L)$ is the Luminosity Function. \citet{efron1992simple} developed a nonparametric technique for estimating $\rho(z)$ and $\psi(L)$ based on the $c^{-}$ method of \citet{lynden1971method}. Luminosity is assumed to have a simple power law redshift dependence:

    \begin{equation}
        \label{eq:lumz}
        L(z) = g(z)L_{0} = (1+z)^{\alpha}L_{0}
    \end{equation}

    \noindent such that the resulting distribution of $L_{0}$, and hence $\psi(L_{0})$ (the local Luminosity Function), becomes independent of redshift.
    \newpar

    Consider the data set seen in Figure \ref{fig:Y15zonea}. Instead of dealing with the entire data set, we deal with subsets that can be constructed independent of the truncation limit that affects the full data set. Each of these subsets includes only the objects for a given range of luminosity and redshift. For the \emph{i}th data point in the $(L,z)$ data set, we can define $J_{i}$ as, 

    \begin{equation}
        \label{eq:Ji}
        J_{i} = \{ j|L_{j} \geqslant L_{i}, z_{j} \leqslant z_{i}^{max} \} ~,
    \end{equation}

    \noindent where $L_{i}$ is the \emph{i}th GRB luminosity and $z_{i}^{max}$ is the maximum redshift at which a GRB with luminosity $L_{i}$ can be observed due to the detector threshold limit. This produces the dashed black bounding box seen in Figure \ref{fig:Y15zonea}. $N_{i}$ represents the number of GRBs in this region: $N_{i} \equiv {\rm size of }\{J_{i}\}$. This is the same as in the $c^{-}$ method in \citet{lynden1971method}. $R_{i}$ is the number of events that have redshift $z_{j}$ less than $z_{i}$, 

    \begin{equation}
        \label{eq:Ri}
        R_{i} \equiv {\rm size of }\{j \in J_{i}: z_{j} < z_{i} \} ~.
    \end{equation}

    \noindent We expect the rank $R_{i}$ of $z_{i}$ to be uniformly distributed between 1 and $N_{i}$ if $L$ and $z$ are independent of each other \citep{efron1992simple}. The Efron-Petrosian test statistic $\tau$ is then,  

    \begin{equation}
        \label{eq:tau}
        \tau \equiv \frac{\sum_{i}(R_{i}-E_{i})}{\sqrt{\sum_{i}V_{i}}} ~,
    \end{equation}

    \noindent where $E_{i} = (1+N_{i})/2$, $V_{i} = (N_{i}^2 - 1)/12$ are the expected mean and the variance of $R_{i}$, respectively. This is a specialized version of Kendell's $\tau$ statistic. The $\tau$ statistic represents the significance of the correlation in the bivariate distribution of the two quantities of interest by taking into account the effects of data truncation created by a detection threshold hard cutoff. It is normally distributed about a mean of 0 with a standard deviation of 1 \citep{efron1992simple}. Hence, a $\tau$ of 2 implies a $2\sigma$ correlation. Once the existence of a correlation has been confirmed, one simply parameterizes it in some way to remove the correlation. As has been aforementioned, a functional form of $\liso$ as $L \rightarrow L_{0} = L/g(z)$ has been chosen in majority of previous studies, where $g(z)$ has a simple power law dependence on $z$, $g(z) = (1+z)^{\alpha}$. Then, it is just a matter of varying $\alpha$ until $\tau = 0$.
    \newpar

    In this work we use the following cosmological parameters: $H_{0} = 71~[~km~s^{-1}~Mpc^{-1}~]$, $\Omega_{m} = 0.3$, $\Omega_{\Lambda} = 0.7$ \citep[e.g.,]{shahmoradi2011possible, shahmoradi2013multivariate, shahmoradi2015short}.

\section{Reanalyses of past work}
\label{sec:analysis}

    In the following subsections, we attempt to regenerate and reanalyze the findings of several recent papers that present evidence in favor of a strong evolution of the GRB luminosity/energetics with redshift.

    \subsection{\citet{yu2015unexpectedly} (\citetalias{yu2015unexpectedly})}
    \label{sec:analysis:Y15}

        \begin{figure*}
            \centering
            \makebox[\textwidth]
            {
                \begin{tabular}{ccc}
                    \subfloat[]{\includegraphics[width=0.31\textwidth]{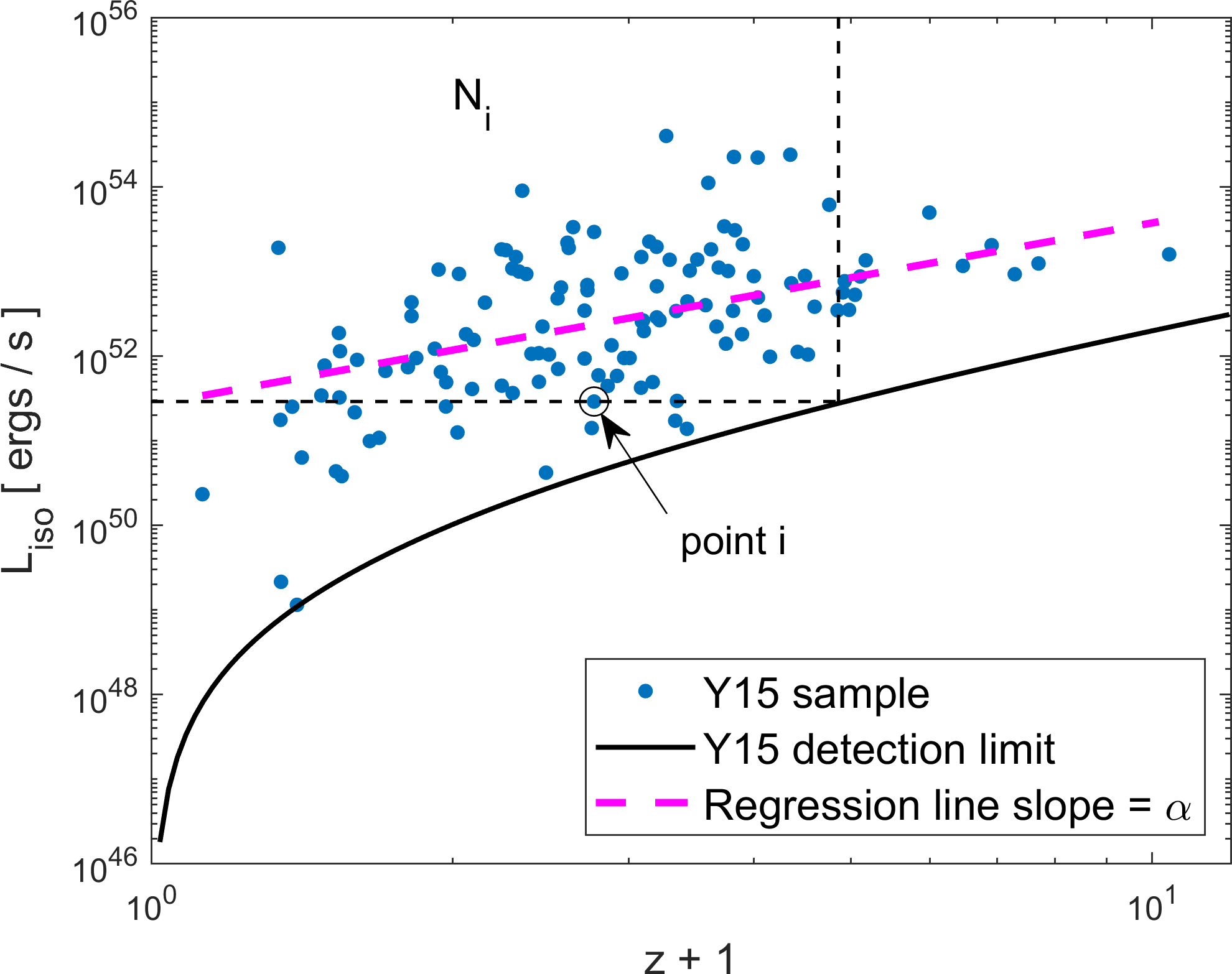} \label{fig:Y15zonea}} &
                    \subfloat[]{\includegraphics[width=0.31\textwidth]{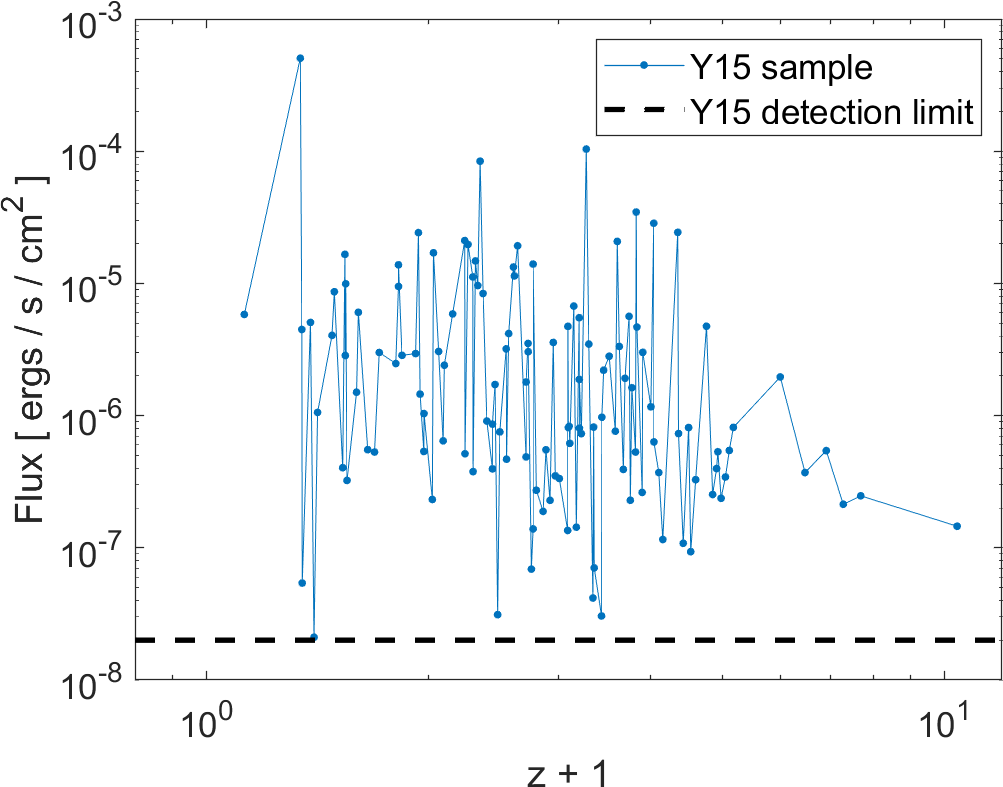} \label{fig:Y15zoneb}} &
                    \subfloat[]{\includegraphics[width=0.31\textwidth]{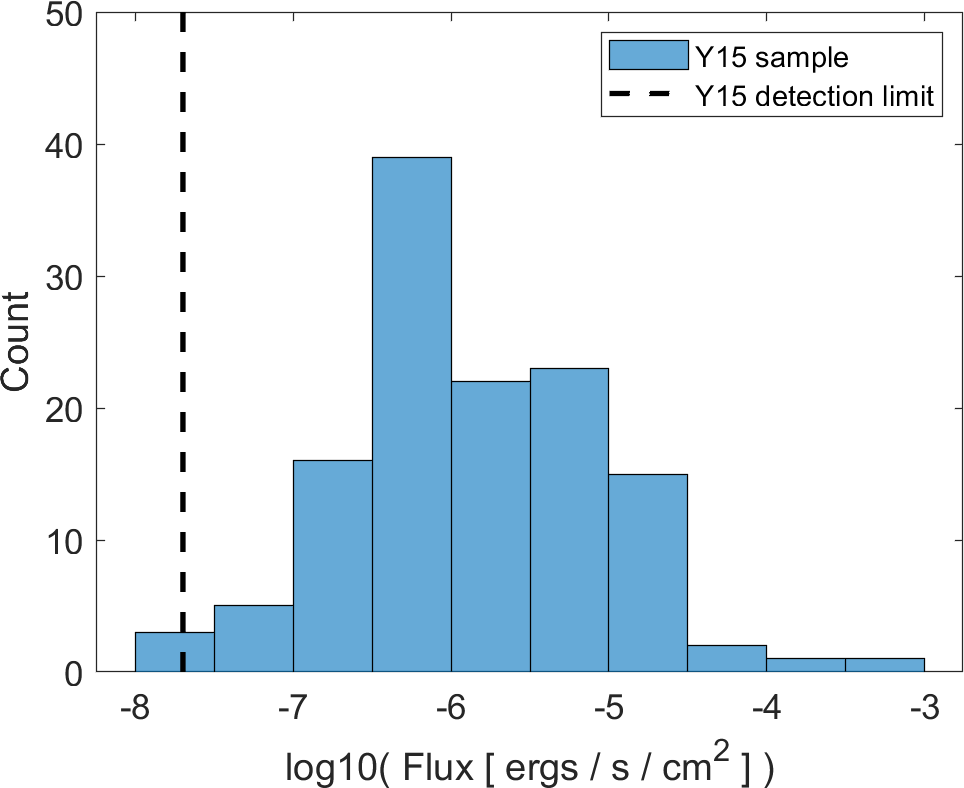} \label{fig:Y15zonec}} \\
                    \subfloat[]{\includegraphics[width=0.31\textwidth]{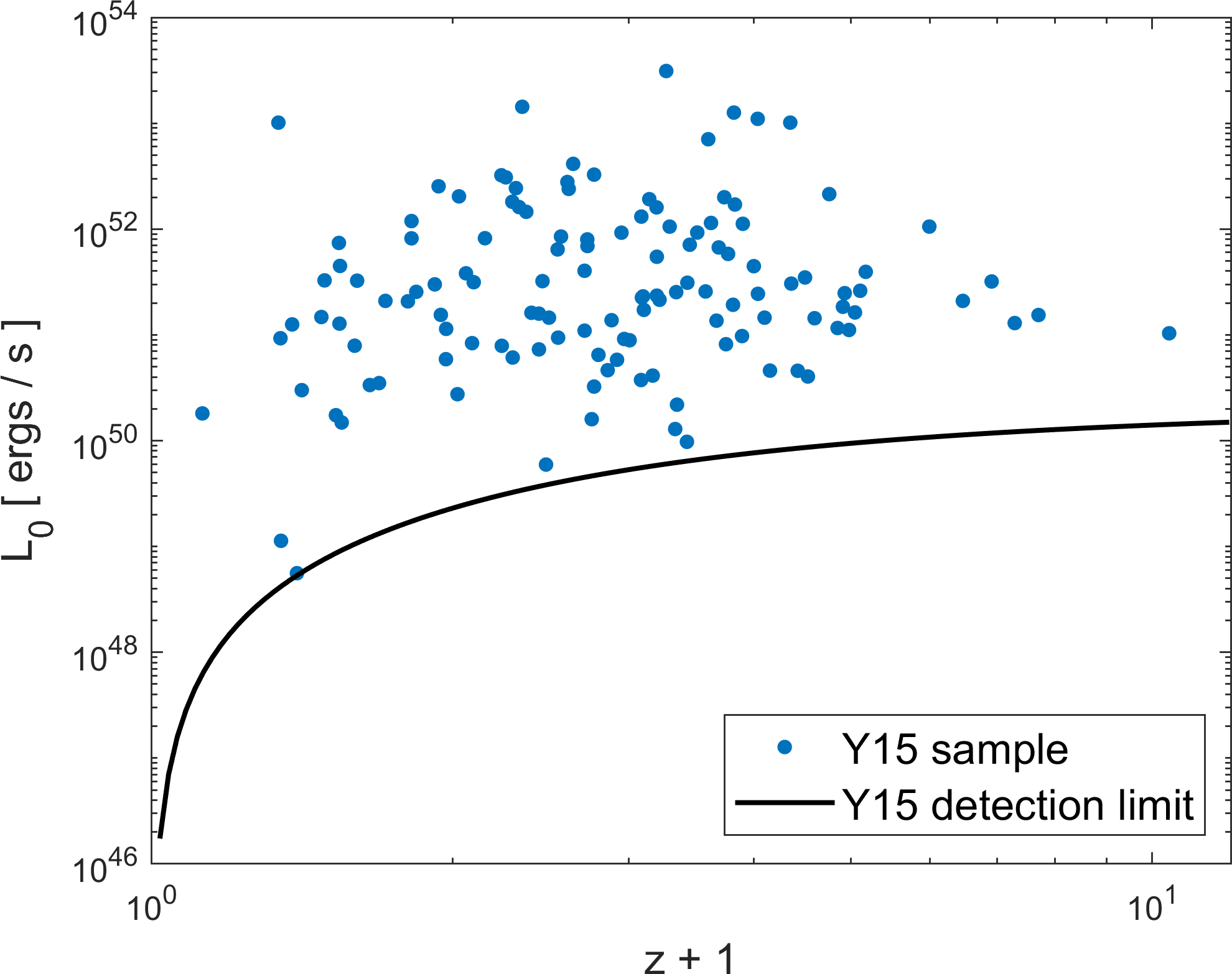} \label{fig:Y15zoned}} &
                    \subfloat[]{\includegraphics[width=0.31\textwidth]{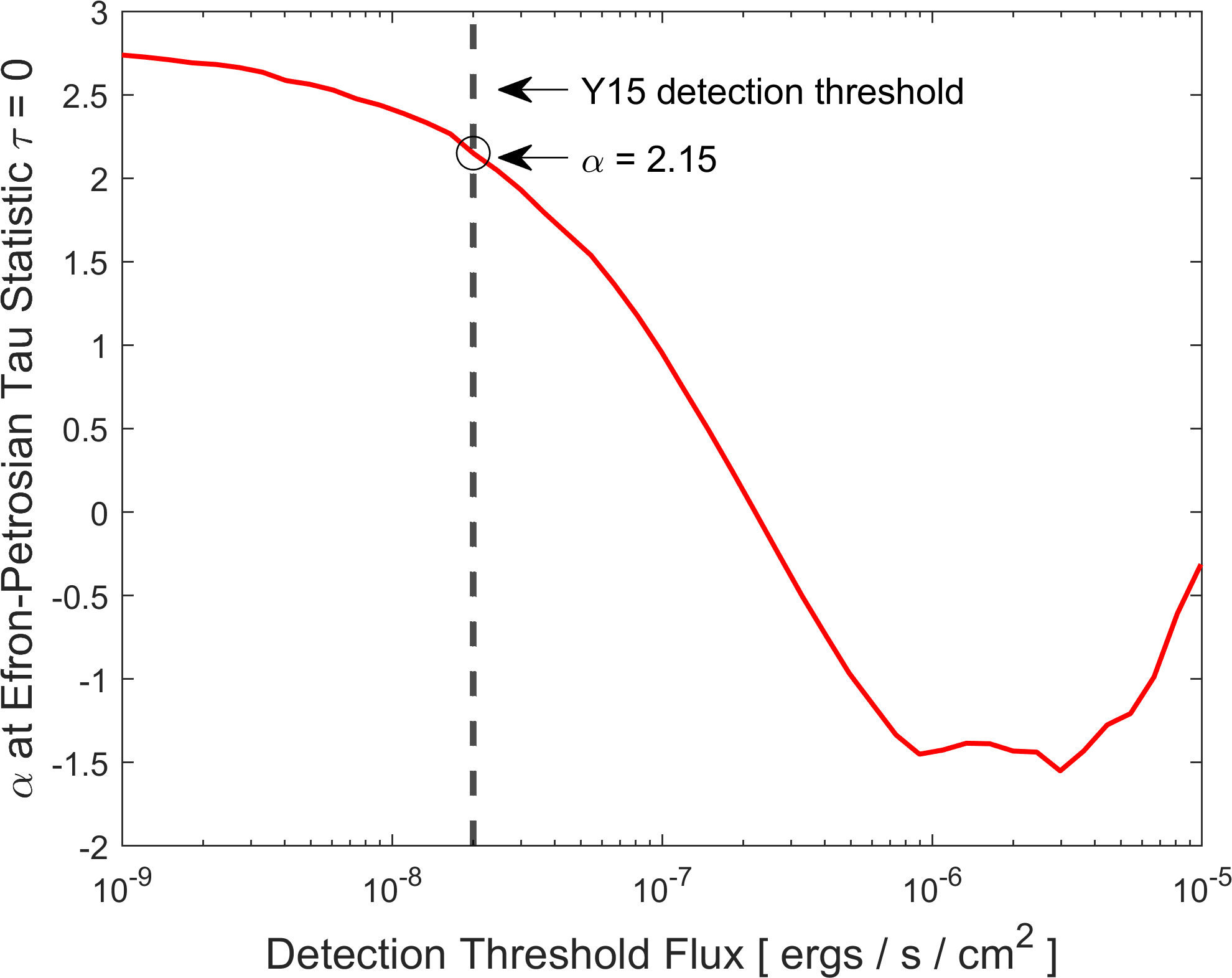} \label{fig:Y15zonee}} &
                    \subfloat[]{\includegraphics[width=0.31\textwidth]{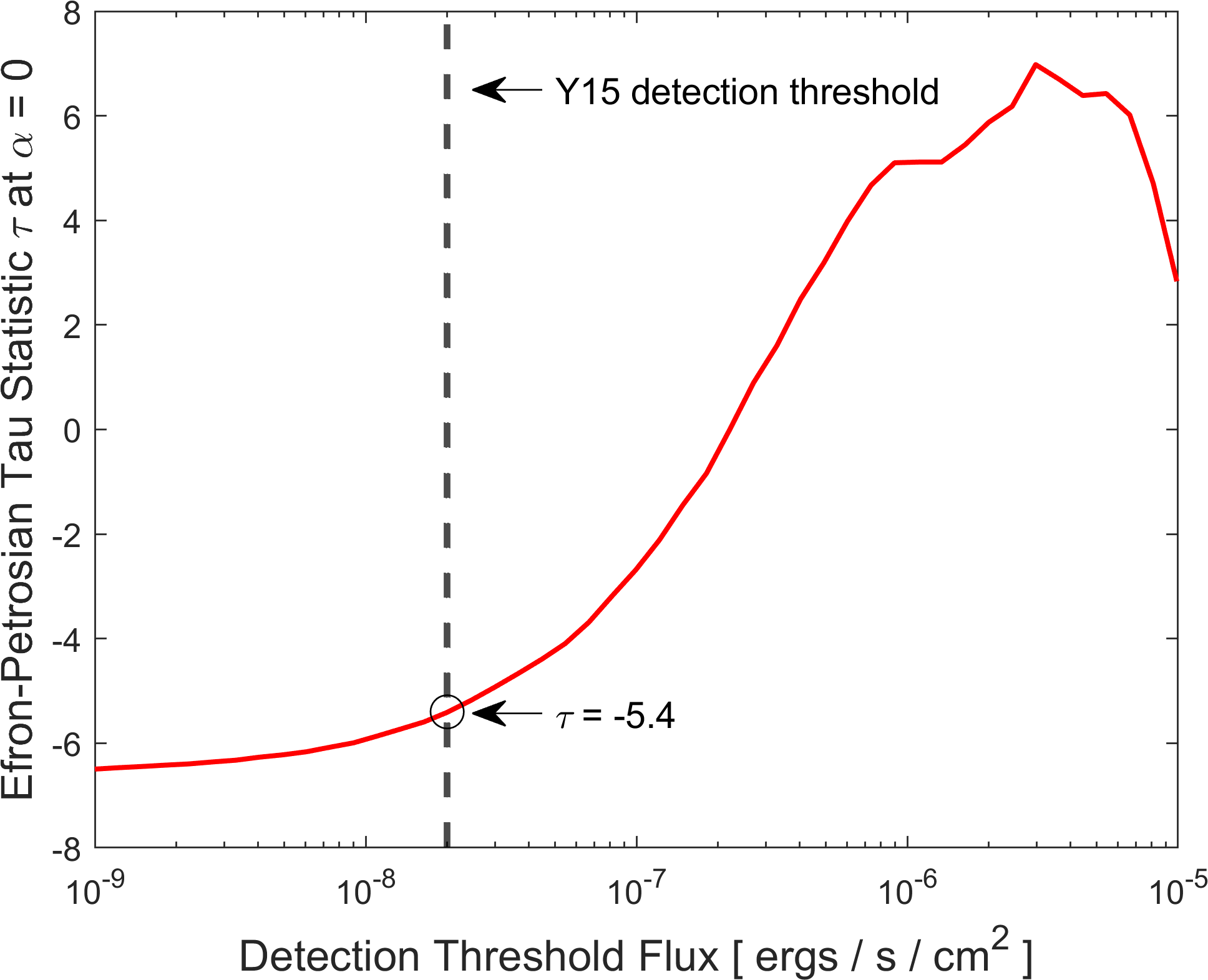} \label{fig:Y15zonef}} \\
                \end{tabular}
            }
            \caption{
            Plotting of the 127 GRBs from \citetalias{yu2015unexpectedly}. Plot (a) shows isotropic luminosity vs. redshift. The black line represents the observational limit of \emph{Swift} assumed by \citetalias{yu2015unexpectedly} to be $2.0 \times 10^{-8}~[~erg~cm^{-2}~s^{-1}~]$. The purple line represents the linear regression through the data set whose slope is $\alpha=2.15$. Plot (b) is the observer frame visualization of the \citetalias{yu2015unexpectedly} data set, where the dashed line is the observational limit of \citetalias{yu2015unexpectedly}. Plot (c) is a histogram of flux, where the dashed line is the observational limit. Plot (d) shows redshift vs. $L_{0}=L(z)/(1+z)^{2.15}$, the redshift-independent luminosity. Plot (e) shows a range of possible threshold limits vs. $\alpha$ values at $\tau = 0$. The intersection of this line with \citetalias{yu2015unexpectedly}'s threshold limit is our value for their $\alpha$. Plot (f) shows a range of possible threshold limits vs $\tau$ values at $\alpha = 0$. Assuming the detection threshold of \citetalias{yu2015unexpectedly}, a redshift-independent luminosity distribution is rejected at $5.4\sigma$. However, choosing a more conservative detection threshold at ~ $2\times10^{-7}~[~erg~cm^{-2}~s^{-1}~]$ yields no evidence for luminosity-redshift evolution.
            \label{fig:Y15zone}
            }
        \end{figure*}

        In their paper, \citetalias{yu2015unexpectedly} find an excess of LGRBs at low redshift ($z<1$), deviating from the SFR. They use the method of \citet{efron1992simple} to infer the luminosity function and the cosmic rates of LGRBs based on the observational data collected by \emph{Swift}'s Burst Alert Telescope (BAT) gamma-ray detector. They find that the luminosity function of LGRBs evolves with redshift as Eq. \eqref{eq:lumz} with $\alpha = 2.43^{+0.41}_{-0.38}$. This conclusion is based on the assumption of a flux lower limit of

        \begin{equation}
            \label{eq:fminY15}
            F_{min} = 2.0 \times 10^{-8}~[~erg~cm^{-2}~s^{-1}~] ~,
        \end{equation}

        \noindent representing the detector threshold limit of \emph{Swift}'s BAT.
        \newpar

        To better understand the role of \emph{Swift}'s BAT detector threshold on the conclusions drawn by \citetalias{yu2015unexpectedly}, here we attempt to reproduce their analysis of \emph{Swift} data. Figures \ref{fig:Y15zonea} - \ref{fig:Y15zonec} depict the distributions of the observational LGRB sample used in Table 1 of \citetalias{yu2015unexpectedly}. Specifically, the bivariate distribution of the 1-second total isotropic peak flux ($\liso$) and the redshifts ($1+z$) of LGRBs as shown in plot \ref{fig:Y15zonea} exhibits an apparently strong correlation. However, much of this correlation is potentially due to the BAT detection threshold effects on the observational sample of LGRBs. To quantify and eliminate the effects of detector threshold, \citetalias{yu2015unexpectedly} use the proposed non-parametric methodology of \citet{efron1992simple} by assuming a parametrization as seen in Eq. \eqref{eq:lumz} for the luminosity-redshift dependence in the LGRB data, such that the resulting distribution of $L_0$ becomes independent of redshift. We note an apparent inconsistency between Eq. \eqref{eq:tau} that we have extracted from \citet{efron1992simple} and Eq. (9) of \citetalias{yu2015unexpectedly}, where summations are taken in different places.
        \newpar

        The value of exponent that we infer from the data set of \citetalias{yu2015unexpectedly} is consistent with, although not the same as, their inferred $\alpha$ exponent. \citetalias{yu2015unexpectedly} found a value of $\alpha = 2.43^{+0.41}_{-0.38}$ in their analysis. In our reanalysis of their work, we find a value of $\alpha = 2.15^{+0.34}_{-0.37}$. However we do believe that their assumption for the value of the flux lower limit of Eq. \eqref{eq:fminY15} is likely a severe underestimation of the detection threshold of \emph{Swift}'s BAT. In reality the detection threshold limit of the BAT detector is far more complex than a simple hard cutoff, but rather, a fuzzy range arising from a multitude of factors.
        \newpar

        The \emph{Swift} satellite is very well known for its immensely complex triggering algorithm. To our knowledge, it is comprised of at least three separate detection mechanisms that complement each other \citep[e.g.,][]{fenimore2003trigger}:

        \begin{enumerate}
            \item
                The first type of trigger is for short time scales (4 ms to 64 ms). These are traditional triggers (single background), for which about 25,000 combinations of time-energy-focal plane subregions are checked per second.
            \item
                The second type of trigger is similar to HETE: fits multiple background regions to remove trends for time scales between 64 ms and 64 s. About 500 combinations for these triggering mechanisms are checked per second. For these rate triggers, false triggers and variable non-GRB sources are also rejected by requiring a new source to be present in an image.
            \item
                The third type of trigger works on longer time scales (minutes) and is based on routine images that are made of the field of view.
        \end{enumerate}

        The entire complexity of the detection mechanism of \emph{Swift}'s BAT, as mentioned above, is summarized in a single value Eq. \eqref{eq:fminY15} in the work of \citetalias{yu2015unexpectedly}. The consequences of choosing this value is most apparent in Figures \ref{fig:Y15zoneb} and \ref{fig:Y15zonec}, where the data set of detected LGRBs virtually ends before reaching the detection threshold line. This implies that the observational LGRB data set is not constrained by the assumed detection threshold of \citetalias{yu2015unexpectedly}, which is counterintuitive. We \emph{do} expect the threshold to soft-truncate the data set, and this truncation likely occurs closer to the central peak of the histogram of Figure \ref{fig:Y15zonec}.
        \newpar

        This severe underestimation of the detection threshold of BAT by \citetalias{yu2015unexpectedly} is very well seen in Figure \ref{fig:Y15zoned} where we plot the redshift-corrected isotropic peak luminosity vs. redshift. The solid black line in this plot represents the redshift-corrected detection threshold. In agreement with our hypothesis in the previous paragraph, we observe that the resulting redshift-corrected detection threshold of \citetalias{yu2015unexpectedly} resembles almost a flat line at high redshifts. This is another indication that the inferred relationship between $(1+z)$ and $\liso$ is likely heavily influenced by the improperly-modeled detection threshold of BAT.
        \newpar

        A range of threshold values can be seen in Figures \ref{fig:Y15zonee} and \ref{fig:Y15zonef}, as well as their effects on $\alpha$ at $\tau=0$ and, on $\tau$ at $\alpha=0$, respectively.
        \newpar

        \begin{figure}
            \centering
            \includegraphics[width=0.47\textwidth]{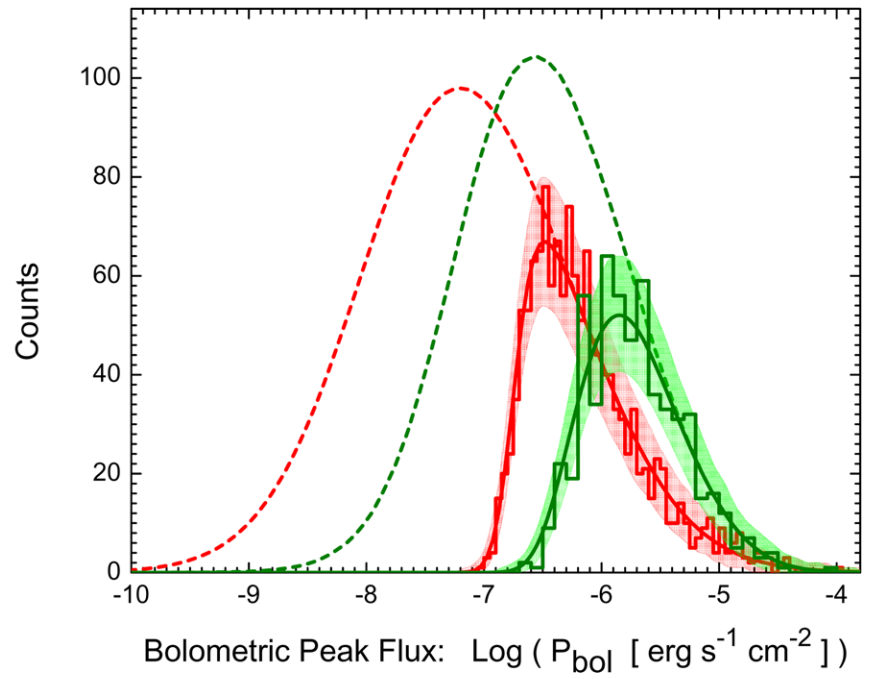}
            \caption{The red and green colors represent data and model for LGRBs and SGRBs, respectively, for BATSE catalogue GRBs. The solid curves illustrate the projection of the multivariate GRB world models of \citet{shahmoradi2013multivariate, shahmoradi2015short} on the BATSE-catalog peak flux $P_{bol}$ distribution, subject to the BATSE detection threshold. The colour-shaded areas represent the 90 percent prediction intervals for the distribution of BATSE data. The dashed lines represent the predicted underlying populations of LGRBs and SGRBs, respectively based on the multivariate GRB world models of \citet{shahmoradi2013multivariate, shahmoradi2015short}.
            \label{fig:Shah15}}
        \end{figure}

        To gain a better insight into the effects of detection threshold on observational data, we have reproduced parts of the results of \citet{shahmoradi2015short} in Figure \ref{fig:Shah15}. This figure illustrates well the subtle fuzzy effects of the BATSE detector threshold on the observed distribution of the energetics of BATSE LGRBs and SGRBs, including the observed peak flux distribution for which a sharp detection threshold cutoff is frequently assumed. The detection threshold causes a deviation in the observed data from the inferred underlying population, to begin just to the right of the histogram peak (the solid lines). This deviation becomes more significant as one moves leftward. \citetalias{yu2015unexpectedly} chose their detector threshold to begin far to the left of the histogram peak, when in reality, it should have been chosen close to the peak.
        \newpar

        Finally, we turn our attention to \citetalias{yu2015unexpectedly}'s Monte Carlo simulations, which seemingly confirms their results. Therein, they begin with their inferred value of $\alpha$ to simulate a distribution of LGRBs following the relationship of Eq. \eqref{eq:lumz} with $\alpha = 2.43$. They find that the synthetic data and the observed data have similar distributions. This is, however, no surprise considering their simulation was based on the derived results of their observational analysis and the assumed potentially-underestimated detection threshold. In other words, their Monte Carlo simulation proves the self-consistency of the Efron-Petrosian statistic and their analysis, but falls short of verifying the accuracy of the detection threshold assumption made in their analysis. Therefore, this circular reasoning concludes that the two observed and synthetic distributions are similar without validating the accuracies of the assumptions made in their work.
        \newpar

        Their primary conclusion of \citetalias{yu2015unexpectedly}, of an excess of GRBs at low redshift ($z<1$) compared to the SFR, also contradicts previous studies based on the properties of GRB host galaxies. In point of fact, \citet{vergani2015long}, \citet{perley2015connecting, perley2016swiftb, perley2016swifta}, \citet{kruhler2015grb}, and \citet{schulze2015optically} performed spectroscopic and multi-wavelength studies on the properties (stellar masses, luminosities, SFR, and metallicity) of GRB host galaxies of various complete GRB samples and compared them to those of the star-forming galaxies selected by galaxy surveys. Their collective results indicate that at low redshift ($z<1$) only a small fraction of the star formation produces GRBs \citet{pescalli2016rate}.

    \subsection{\citet{pescalli2016rate} (\citetalias{pescalli2016rate})}
    \label{sec:analysis:P16}

        \begin{figure*}
            \centering
            \makebox[\textwidth]
            {
                \begin{tabular}{ccc}
                    \subfloat[]{\includegraphics[width=0.31\textwidth]{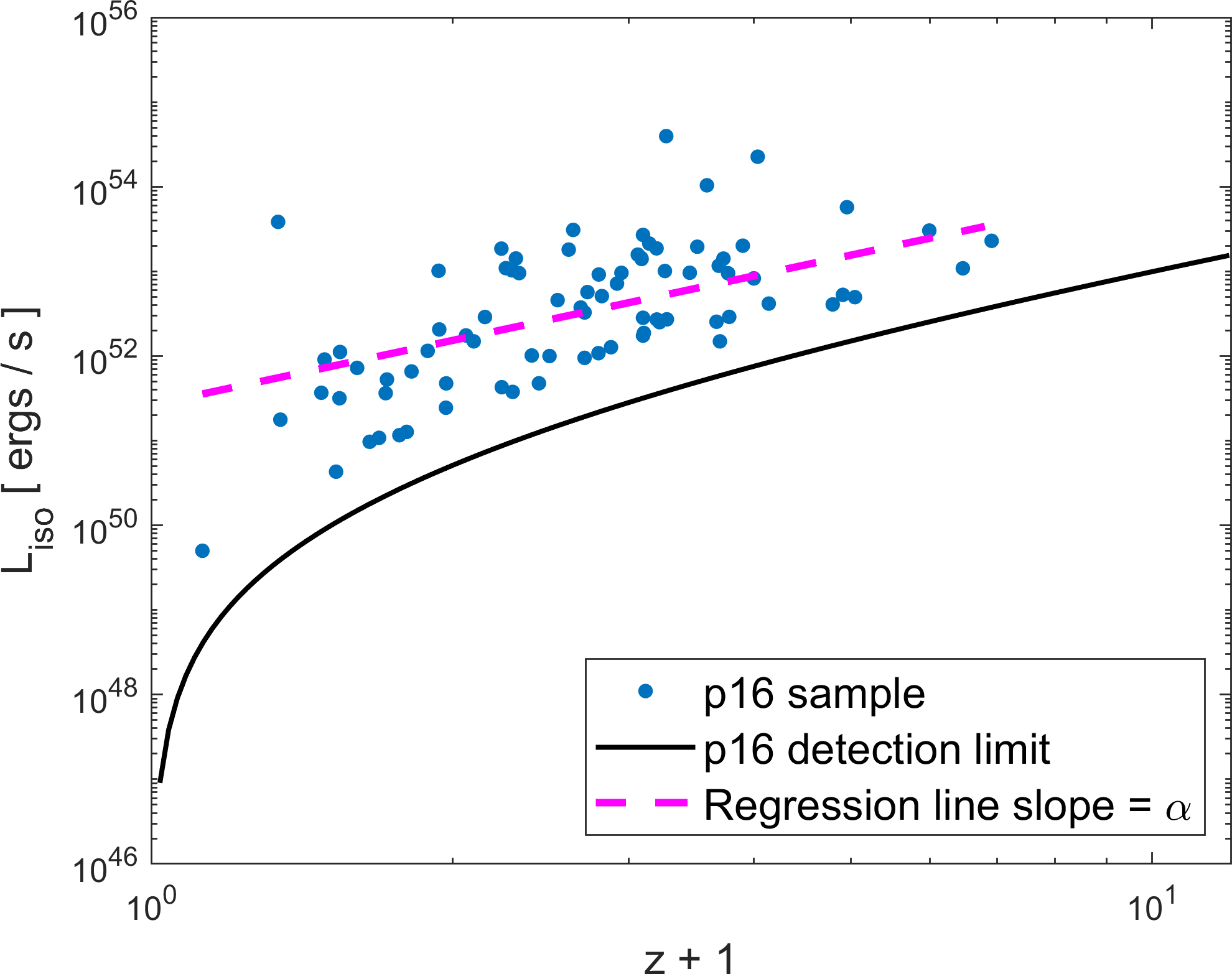} \label{fig:P16figsa}} &
                    \subfloat[]{\includegraphics[width=0.31\textwidth]{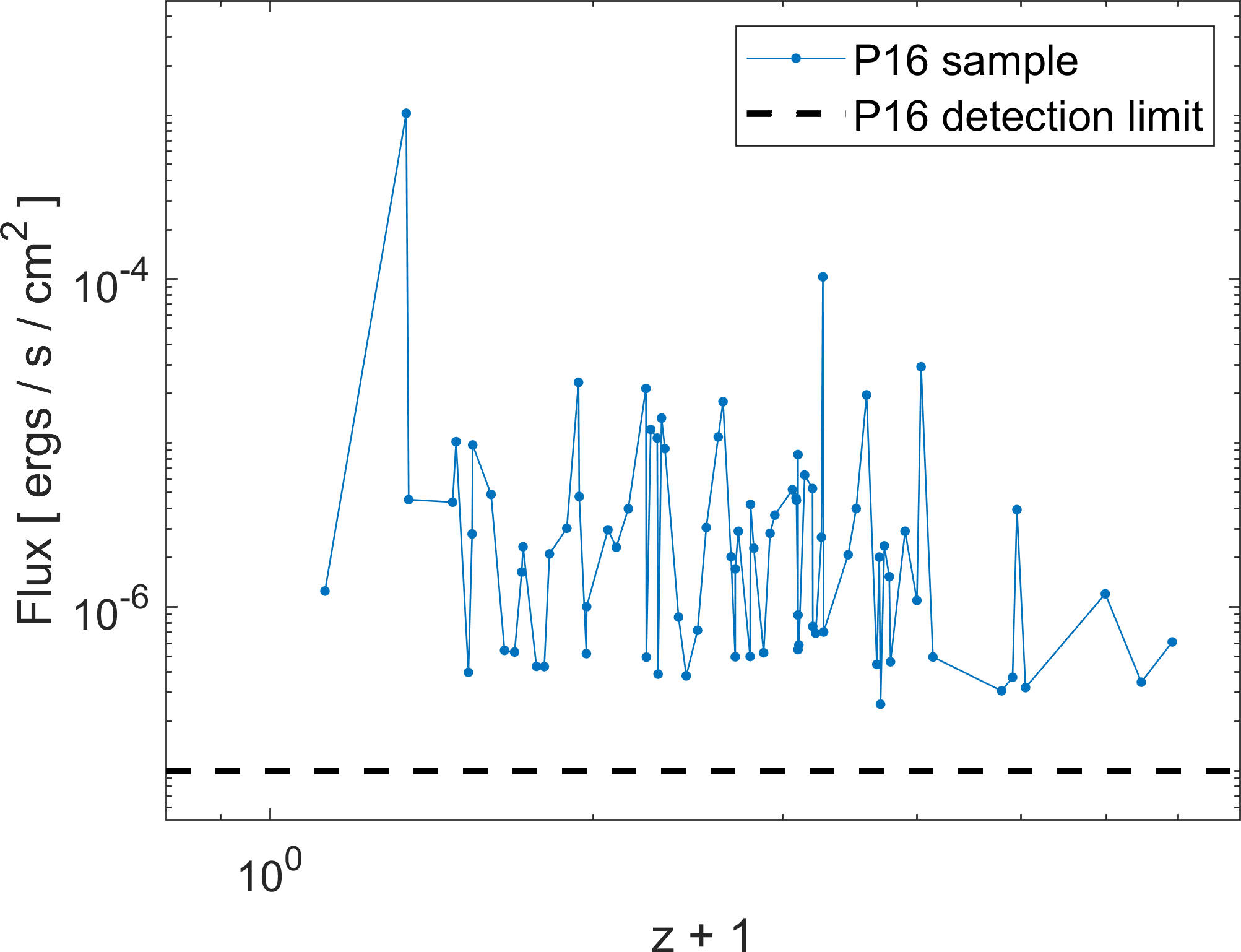} \label{fig:P16figsb}} &
                    \subfloat[]{\includegraphics[width=0.31\textwidth]{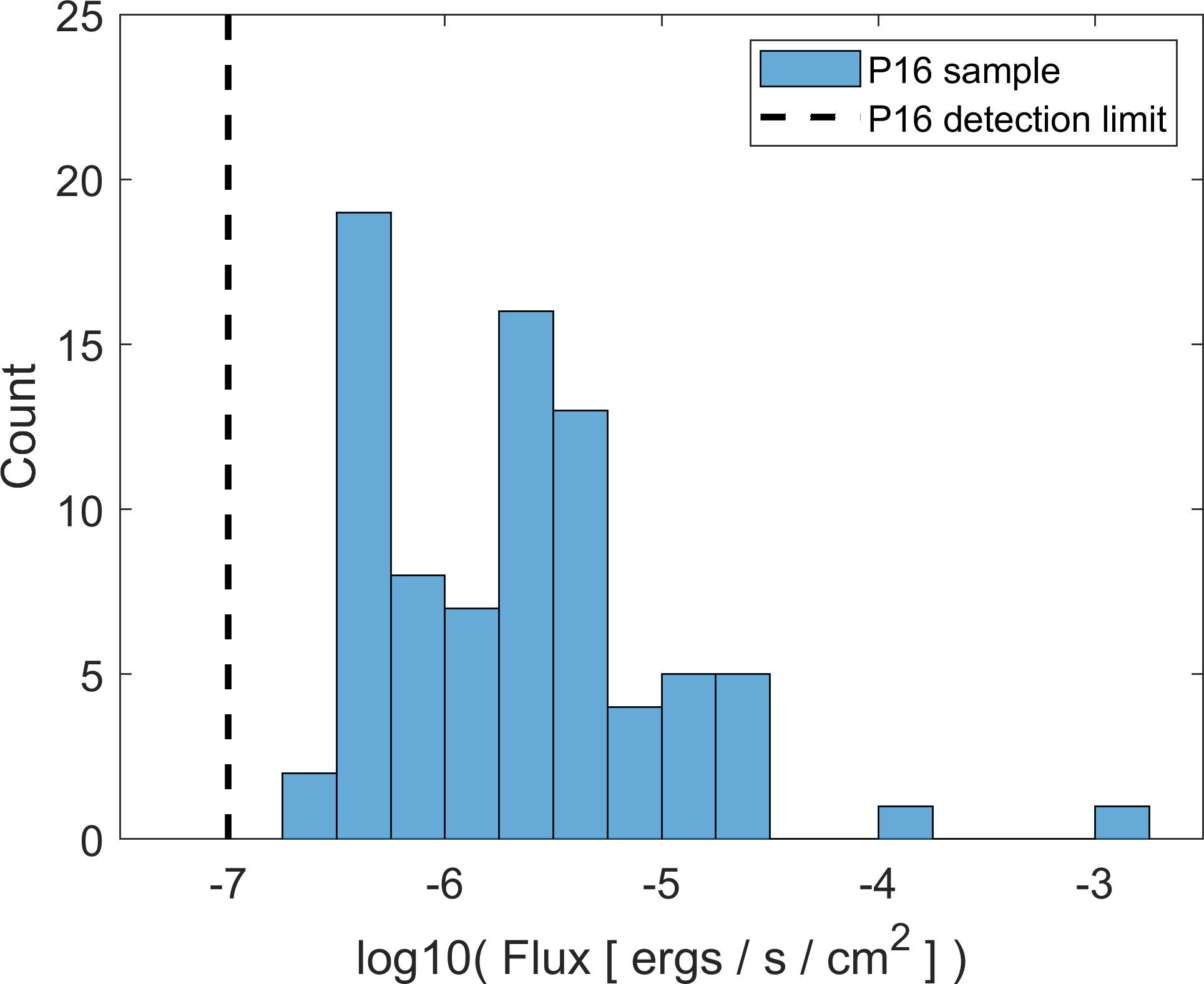} \label{fig:P16figsc}} \\
                    \subfloat[]{\includegraphics[width=0.31\textwidth]{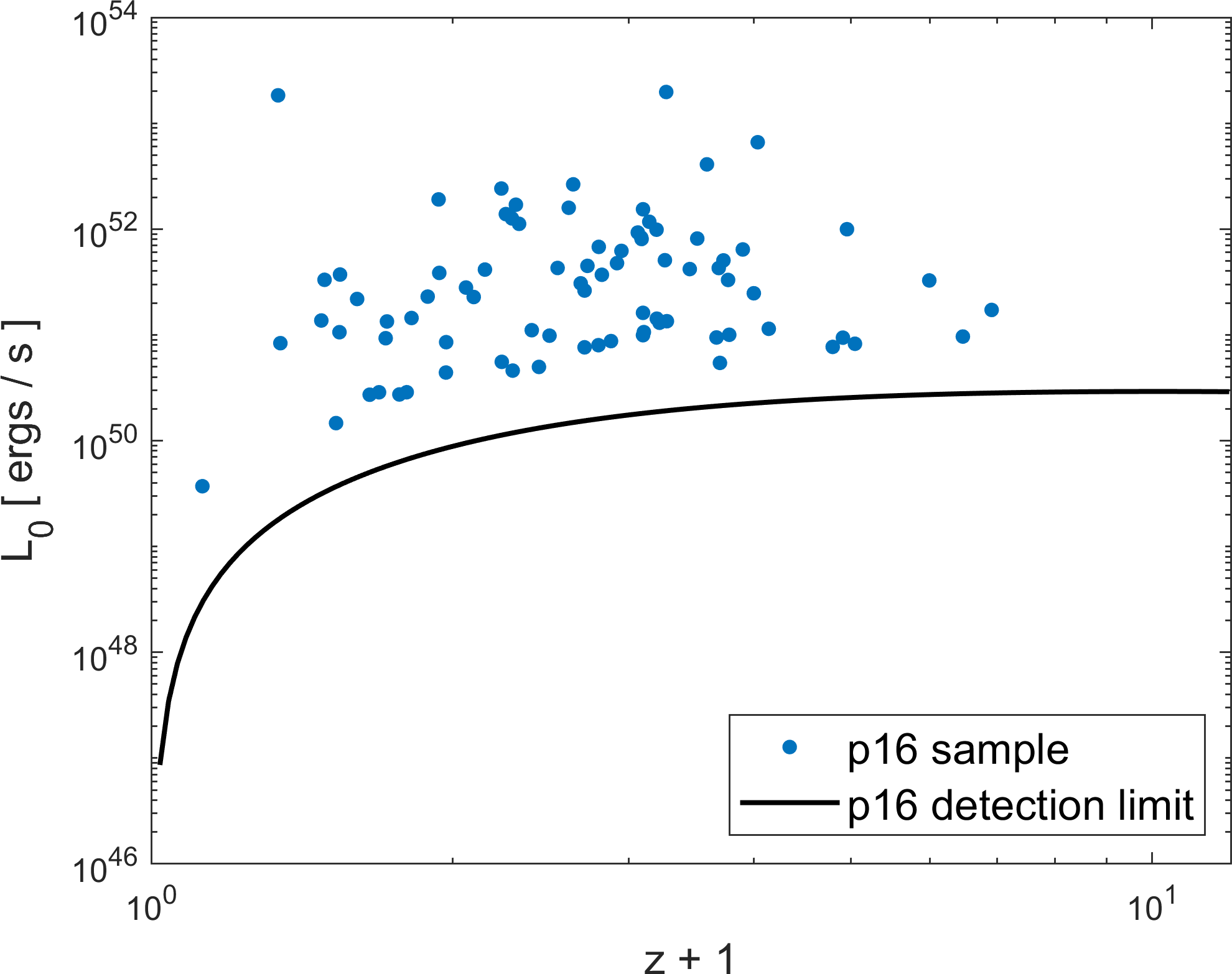} \label{fig:P16figsd}} &
                    \subfloat[]{\includegraphics[width=0.31\textwidth]{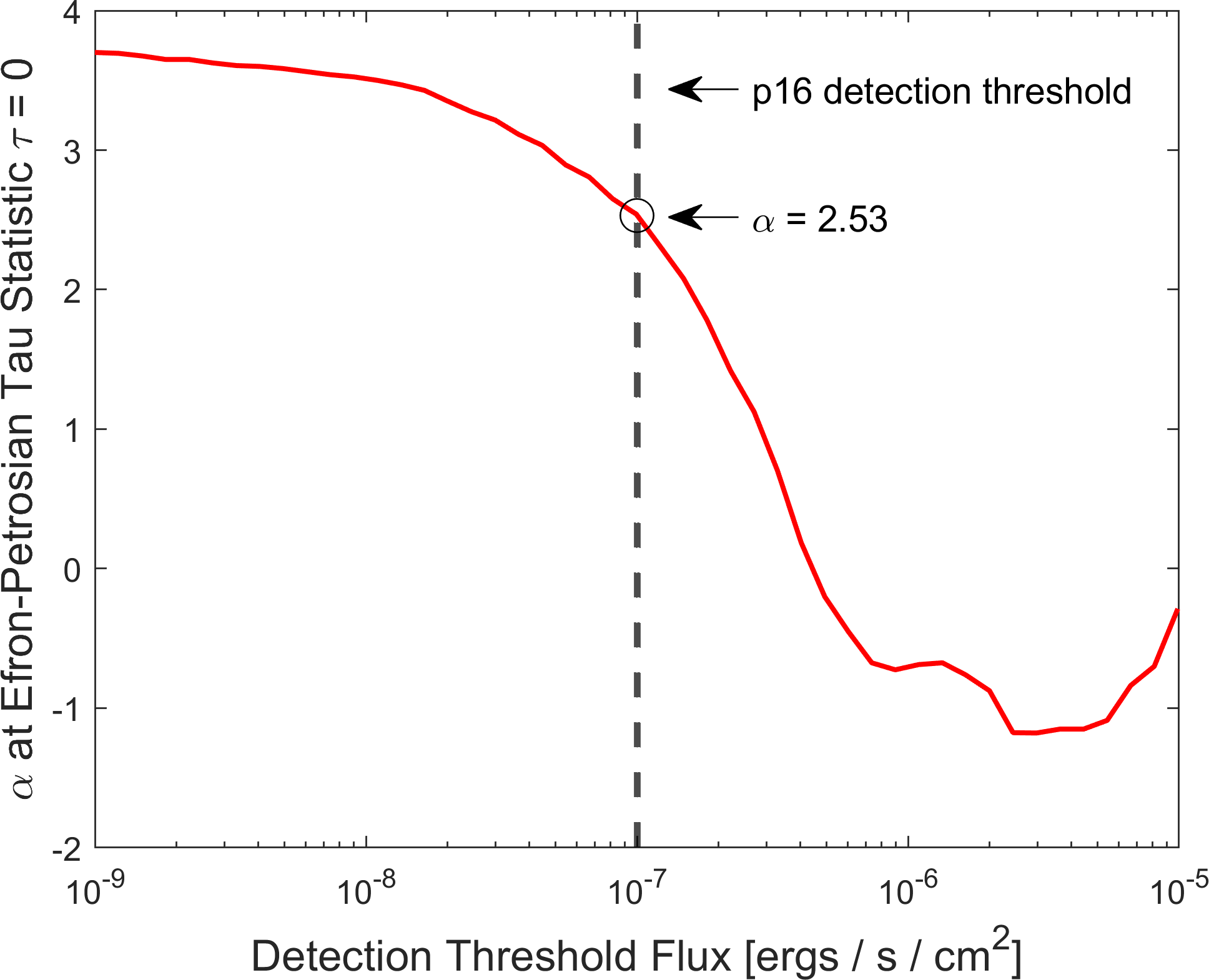} \label{fig:P16figse}} &
                    \subfloat[]{\includegraphics[width=0.31\textwidth]{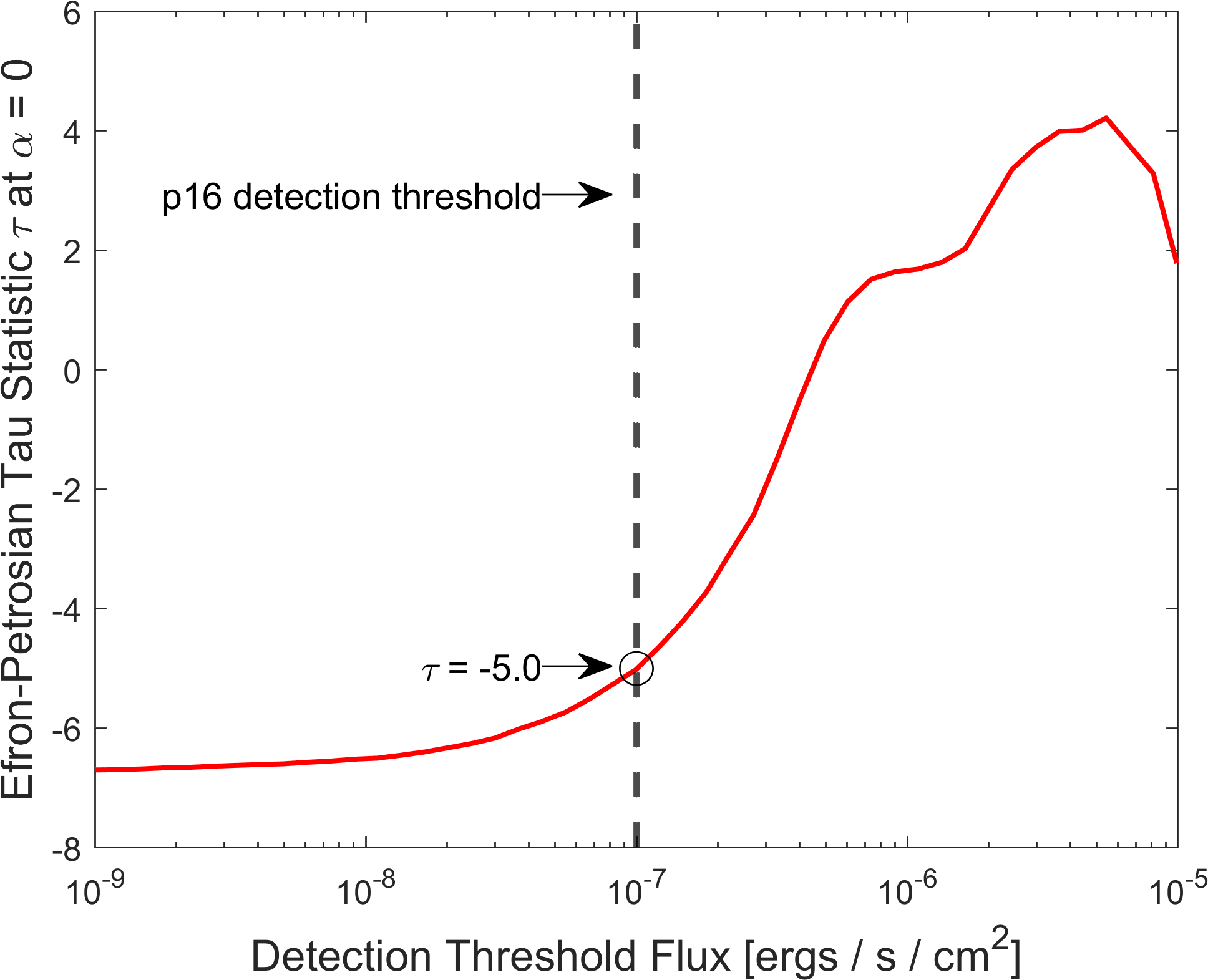} \label{fig:P16figsf}} \\
                \end{tabular}
            }
            \caption{
            Plotting of the 81 GRBs from \citetalias{pescalli2016rate}. Plot (a) shows redshift vs. isotropic luminosity. The black line represents the observational limit of \emph{Swift}, which has been deduced from \citetalias{pescalli2016rate} to be $1.0 \times 10^{-7}~[~erg~cm^{-2}~s^{-1}~]$. The purple line represents the linear regression through the data set whose slope is $\alpha=2.53$. Plot (b) is the observer frame visualization of the \citetalias{pescalli2016rate} data set, where the dashed line is the observational limit. Plot (c) is a histogram of flux, where the dashed line is the observational limit. Plot (d) shows redshift vs. $L_{0}=L(z)/(1+z)^{2.53}$, the redshift-independent luminosity. Plot (e) shows a range of possible threshold limits vs. $\alpha$ values at $\tau = 0$. The intersection of this line with \citetalias{pescalli2016rate}'s supposed threshold limit is our value for their $\alpha$. Plot (f) shows a range of possible threshold limits vs $\tau$ values at $\alpha = 0$. Assuming the detection threshold inferred from the analysis of \citetalias{pescalli2016rate}, a redshift-independent luminosity distribution is rejected at $5\sigma$. However, choosing a more conservative detection threshold at ~ $4\times10^{-7}~[~erg~cm^{-2}~s^{-1}~]$ yields no evidence for luminosity-redshift evolution.
            \label{fig:P16figs}
            }
        \end{figure*}

        \citetalias{pescalli2016rate} proceeds in a similar fashion to \citetalias{yu2015unexpectedly}, beginning with the observational data set of LGRBs found in their Table B.1. We extract from this data set 81 LGRBs that have both redshift and isotropic peak luminosity ($\liso$) values for our reanalysis. They use the Efron-Petrosian $\tau$ statistic to find an $\alpha$ value of $\alpha = 2.5$, consistent with the results of \citetalias{yu2015unexpectedly}. This provides the functional form of the luminosity evolution with redshift $L(z) = L_{0}(1+z)^{2.5}$. From here they proceed with Lynden-Bell’s $c^{-}$ method to derive the cumulative luminosity function $\Phi(L_{0})$ and the LGRB formation rate $\rho(z)$.
        \newpar

        These results, however, are predicated based on a value of detector threshold which \citetalias{pescalli2016rate} gives as,

        \begin{equation}
            \label{eq:pminP16}
            P_{lim} = 2.6~[~photons~cm^{-2}~s^{-1}~] ~.
        \end{equation}

        \noindent This corresponds to an instrument that is $\sim$6 times less sensitive than \emph{Swift}'s BAT \citep{salvaterra2012complete}. In their work,  \citetalias{pescalli2016rate} adopt a slightly different approach to modeling the flux limit of their sample. The quantities $L_{lim}$ and $z_{max}$ that are used in the Efron-Petrosian statistic are computed by adopting individual spectral and temporal properties of LGRBs and applying the corresponding K-corrections. This approach results in a small scatter in the energy flux-limit of their $\liso-(z+1)$ plane. However, they find that this non-uniqueness of the detection threshold has a very small impact in the computation of their $\tau$ statistic.
        \newpar

        Given the lack of sufficient details about the approach proposed by \citetalias{pescalli2016rate} and the fact they find almost no difference between the traditional approach to computing the Efron-Petrosian statistic and their proposed method, here we follow the traditional formal technique for computing the $\tau$ statistic.
        However, since \citetalias{pescalli2016rate} do not provide an effective energy flux limit for their sample in units of $[~erg~cm^{-2}~s^{-1}~]$, we searched for an effective energy-flux detection threshold that would yield a $\tau$ statistic comparable to what \citetalias{pescalli2016rate} find.
        \newpar

        We find that an effective detection threshold of $F_{min} = 1.0 \times 10^{-7}~[~erg~cm^{-2}~s^{-1}~]$ yields a regression slope of $\liso-(z+1)$ correlation of $\alpha=2.53$ which is very close to the reported value by \citetalias{pescalli2016rate}. Alternatively, we also compute the corresponding detection threshold energy flux by converting the photon flux threshold of Eq. \eqref{eq:pminP16} to an effective energy flux limit by assuming a Band model \citep{band1993batse} of the form,

        \begin{equation}
            \label{eq:Band}
            \Phi (E) \propto
            \begin{cases}
                E^{\alpha_{ph}}~ \operatorname{e}^{\big(-\frac{(2+\alpha_{ph})E}{\epk}\big)} & \text{if $E\le\big(\epk\big)\big(\frac{\alpha_{ph}-\beta_{ph}}{2+\alpha_{ph}}\big)$,} \\
            	E^{\beta_{ph}} & \text{if otherwise.}
            \end{cases}
        \end{equation}

        \noindent with low- and high-energy spectral indices of $\alpha_{ph}=-1$ and $\beta_{ph}=-2.25$, respectively, taken from \citet{salvaterra2012complete} and an effective spectral peak energy fixed to the average observed spectral peak energy of the LGRB sample of \citetalias{pescalli2016rate} ($E_{p}=574~keV$). Using this approach, we obtain an effective energy flux limit ($F_{min} = 2.34 \times 10^{-7}~[~erg~cm^{-2}~s^{-1}~]$) for the LGRB sample that results in a completely different value, $\alpha = 1.35$, for the regression slope of the $\liso-(z+1)$ relation than what is reported by \citetalias{pescalli2016rate}.
        \newpar

        Therefore, we conclude that the effective detection threshold ($F_{min} = 1.0 \times 10^{-7}~[~erg~cm^{-2}~s^{-1}~]$) that we previously inferred directly from the Efron-Petrosian statistic should likely resemble more the flux limit that is used but not clearly discussed or shown in the work of \citetalias{pescalli2016rate}. However, the trade off in choosing this value is that the limit appears to be underestimated, as can be seen in Figures \ref{fig:P16figsa} - \ref{fig:P16figsc}. It is not possible to choose a value of $E_{p}$ that both yields an $\alpha$ value in agreement with \citetalias{pescalli2016rate} \emph{and} does not appear to underestimate the detector threshold flux limit.
        \newpar

        Assuming this chosen value for $F_{min}$ is indeed an appropriate approximation for that used by \citetalias{pescalli2016rate}, we are again faced with an underestimation of the true effective value of the detection threshold of Swift, similar to \citetalias{yu2015unexpectedly}. This can be clearly seen in Figure \ref{fig:P16figsd} where we plot the redshift-corrected isotropic peak luminosity vs. redshift, and the solid black line represents the redshift-corrected detection threshold. We observe that this threshold resembles almost a flat line at high redshifts, indicating that the inferred relationship between $(1+z)$ and $\liso$ is likely heavily influenced by the improperly-modeled detection threshold of BAT. A range of threshold values can be seen in Figures \ref{fig:P16figse} and \ref{fig:P16figsf}, as well as their effect on $\alpha$ at $\tau=0$ and $\tau$ at $\alpha=0$, respectively.
        \newpar

        In addition to potential underestimation of the detection threshold, the observational data in the work of \citetalias{pescalli2016rate} also appears to not have been homogeneously collected. Looking at Figure \ref{fig:P16figsc}, the histogram of data appears to be multimodal, implying the presence of some, yet-unknown, selection effects in the process of constructing this data set.
        \newpar

        Finally, we turn our attention to \citetalias{pescalli2016rate}'s Monte Carlo simulation, which seems to confirm their results. Unlike \citetalias{yu2015unexpectedly}, \citetalias{pescalli2016rate} avoid a circular logic inference in their simulations by assuming different $F_{min}$ ($5 \times 10^{-8}~[~erg~cm^{-2}~s^{-1}~]$) and $\alpha$ (2.2) values from their methodology and results. They are able to successfully recover the GRB formation rate and luminosity function that they adopted for their simulated sample.
        \newpar

        They further test the consequences of sample incompleteness in two approaches. In the first approach they randomly remove a fraction of the bursts close to $F_{min}$. In the second approach, they lower $F_{min}$ by a factor of 5, creating an underestimation of its value. Both approaches artificially create sample incompleteness. The result of both realizations of sample incompleteness is to flatten out the GRB formation rate at low redshift, creating the illusion of an excess of low-redshift GRBs relative to the SFR. This result contradicts the simulations and findings of \citetalias{yu2015unexpectedly} and corroborates our conclusions in \S\ref{sec:analysis:Y15}.

    \subsection{\citet{tsvetkova2017konus} (\citetalias{tsvetkova2017konus})}
    \label{sec:analysis:T17}

        \begin{figure*}
            \centering
            \makebox[\textwidth]
            {
                \begin{tabular}{ccc}
                    \subfloat[]{\includegraphics[width=0.31\textwidth]{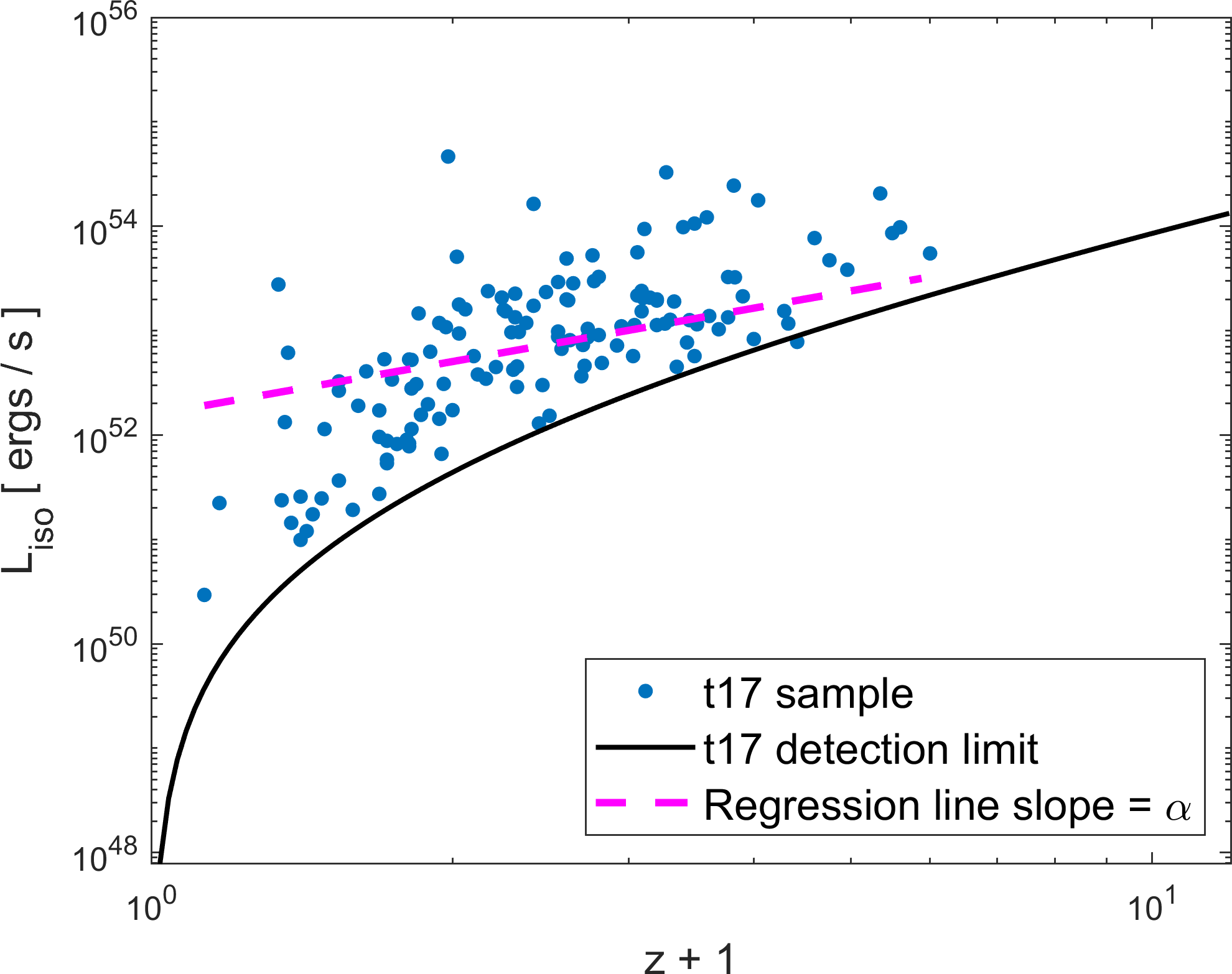} \label{fig:t17figsa}} &
                    \subfloat[]{\includegraphics[width=0.31\textwidth]{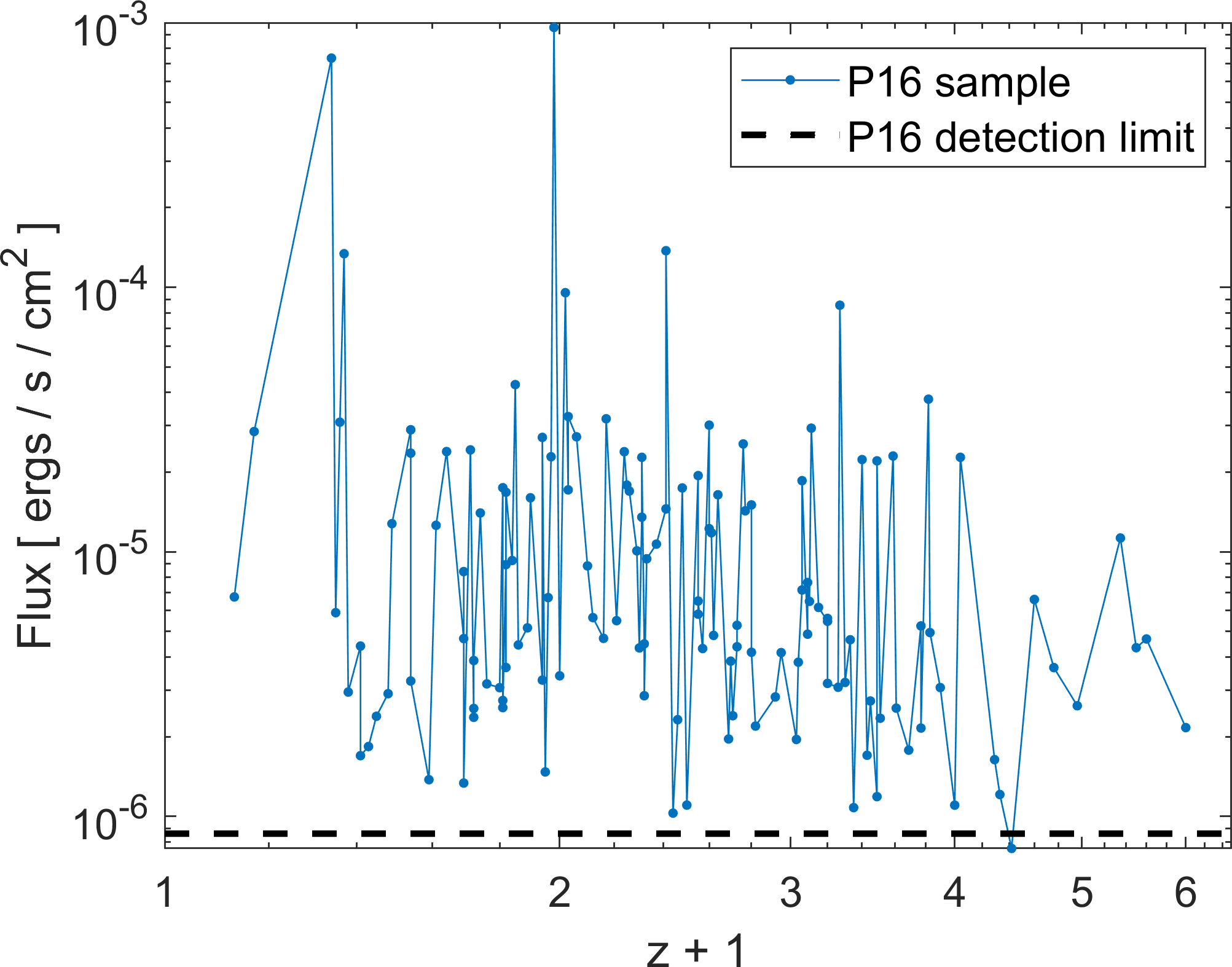} \label{fig:t17figsb}} &
                    \subfloat[]{\includegraphics[width=0.31\textwidth]{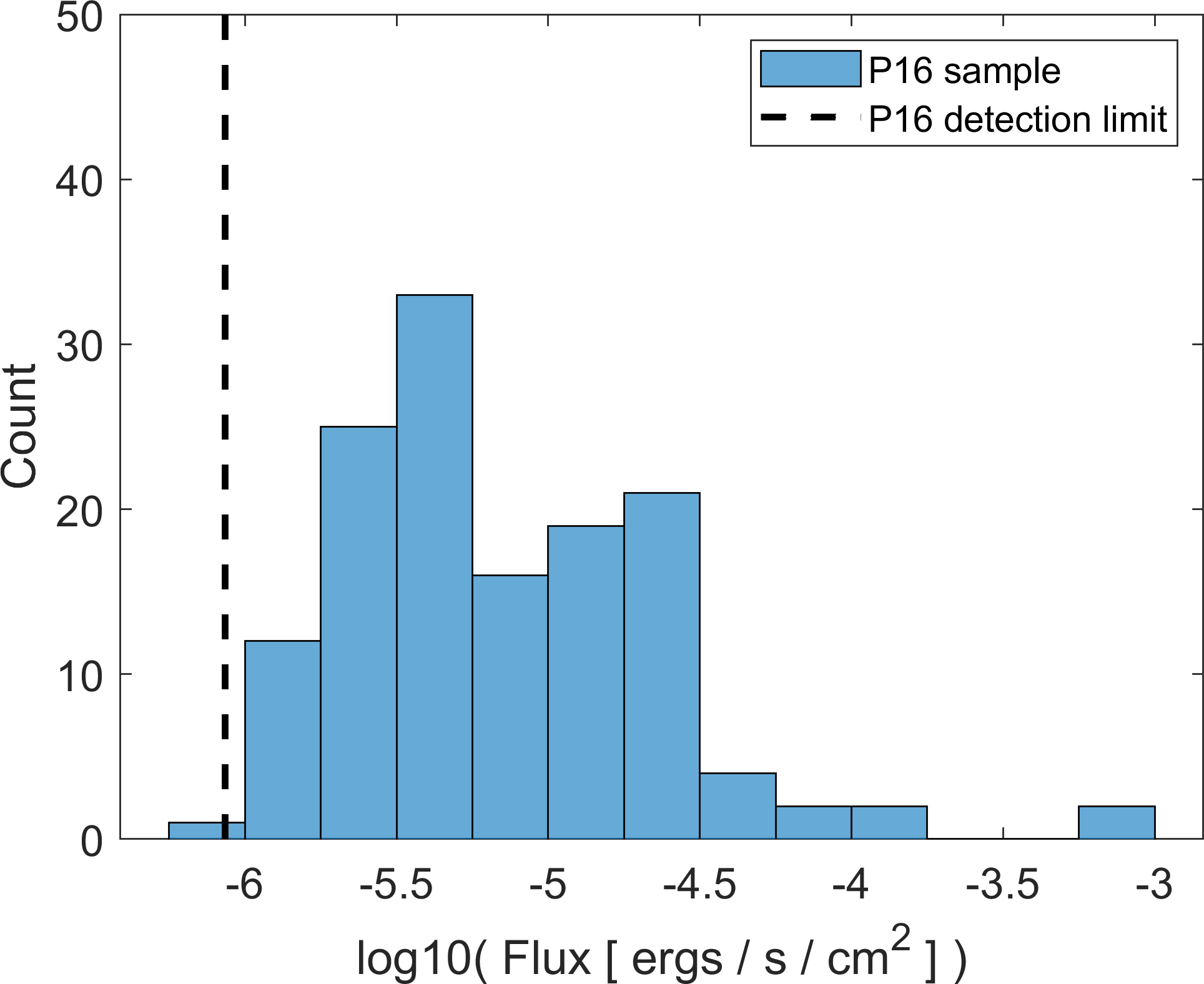} \label{fig:t17figsc}} \\
                    \subfloat[]{\includegraphics[width=0.31\textwidth]{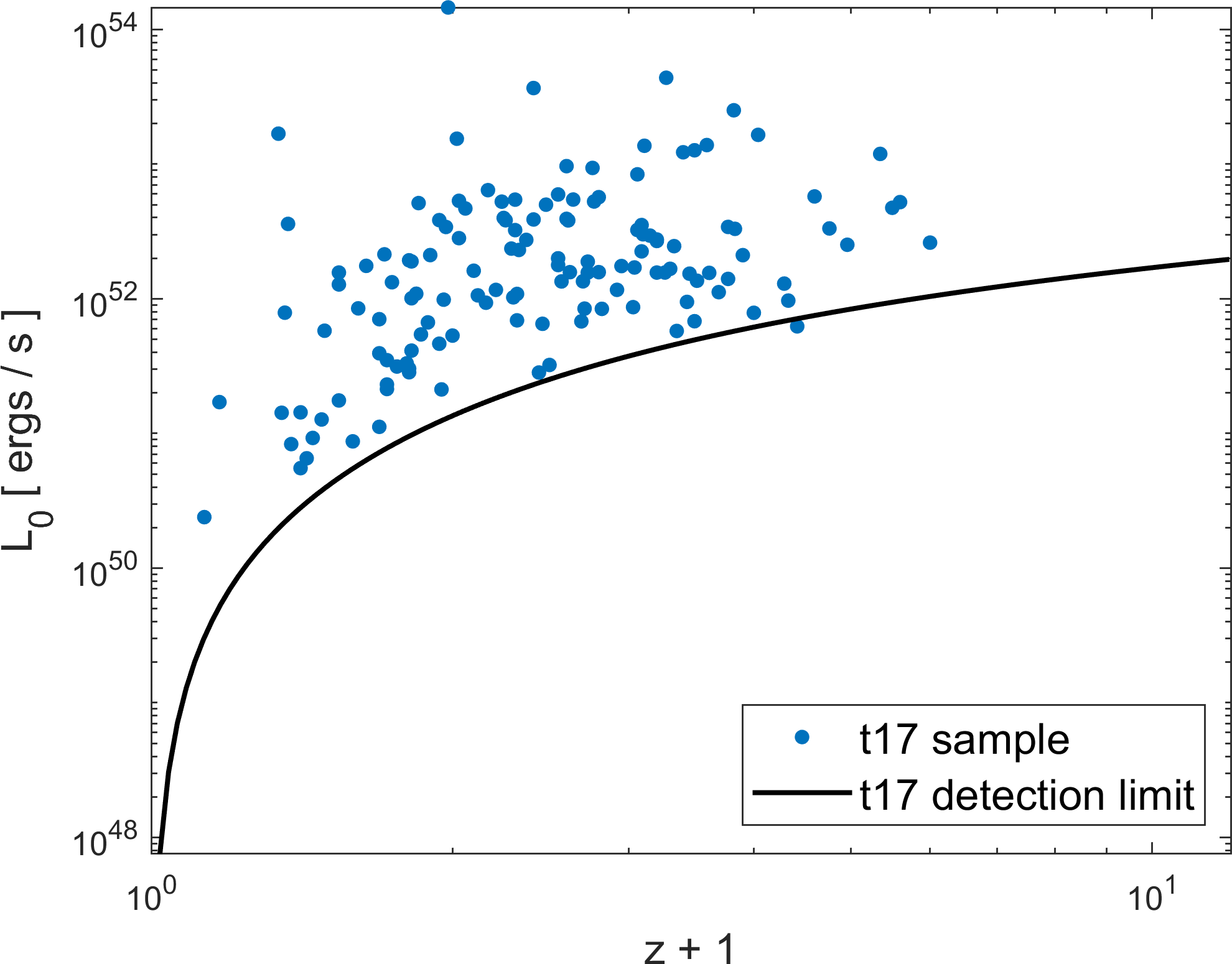} \label{fig:t17figsd}} &
                    \subfloat[]{\includegraphics[width=0.31\textwidth]{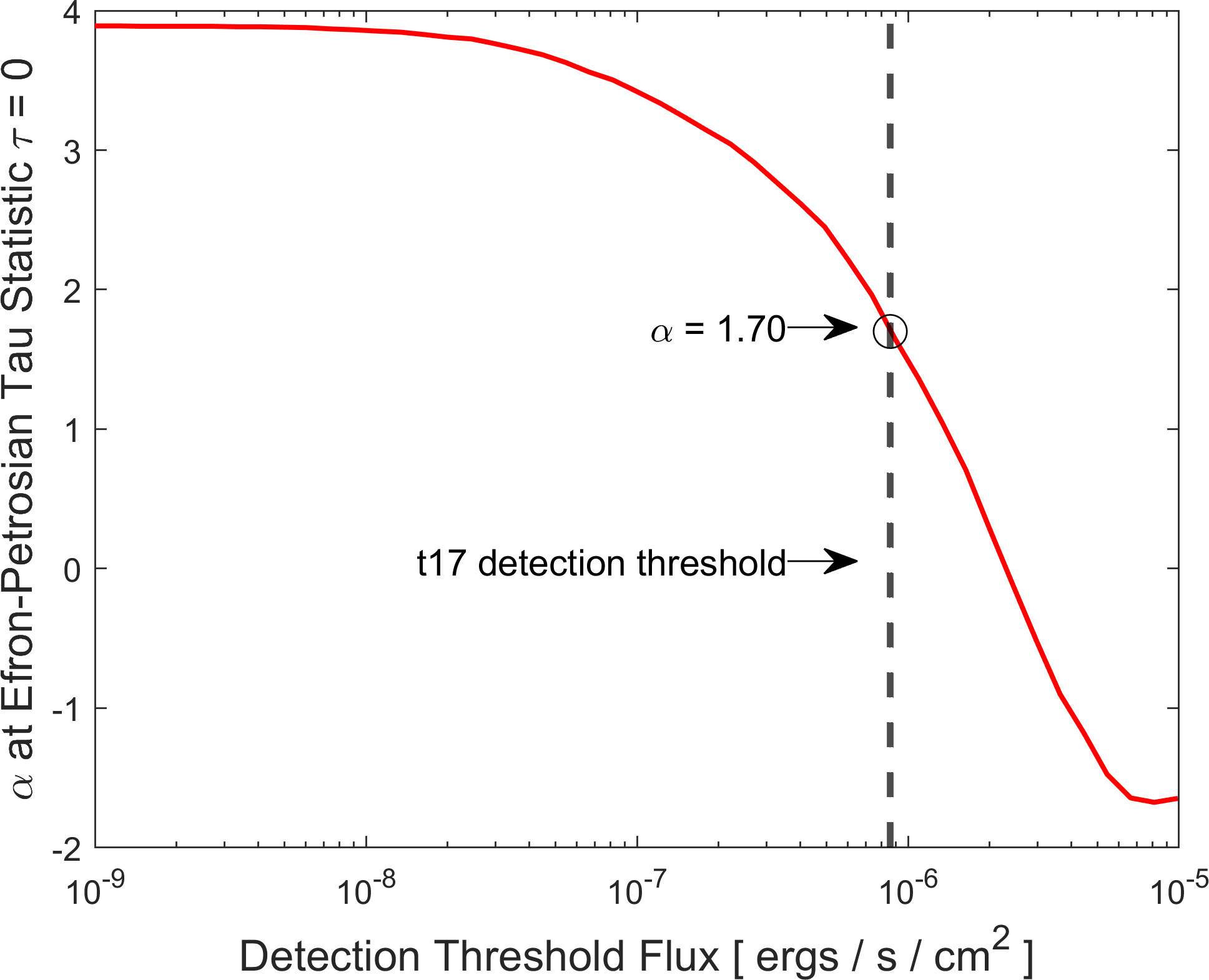} \label{fig:t17figse}} &
                    \subfloat[]{\includegraphics[width=0.31\textwidth]{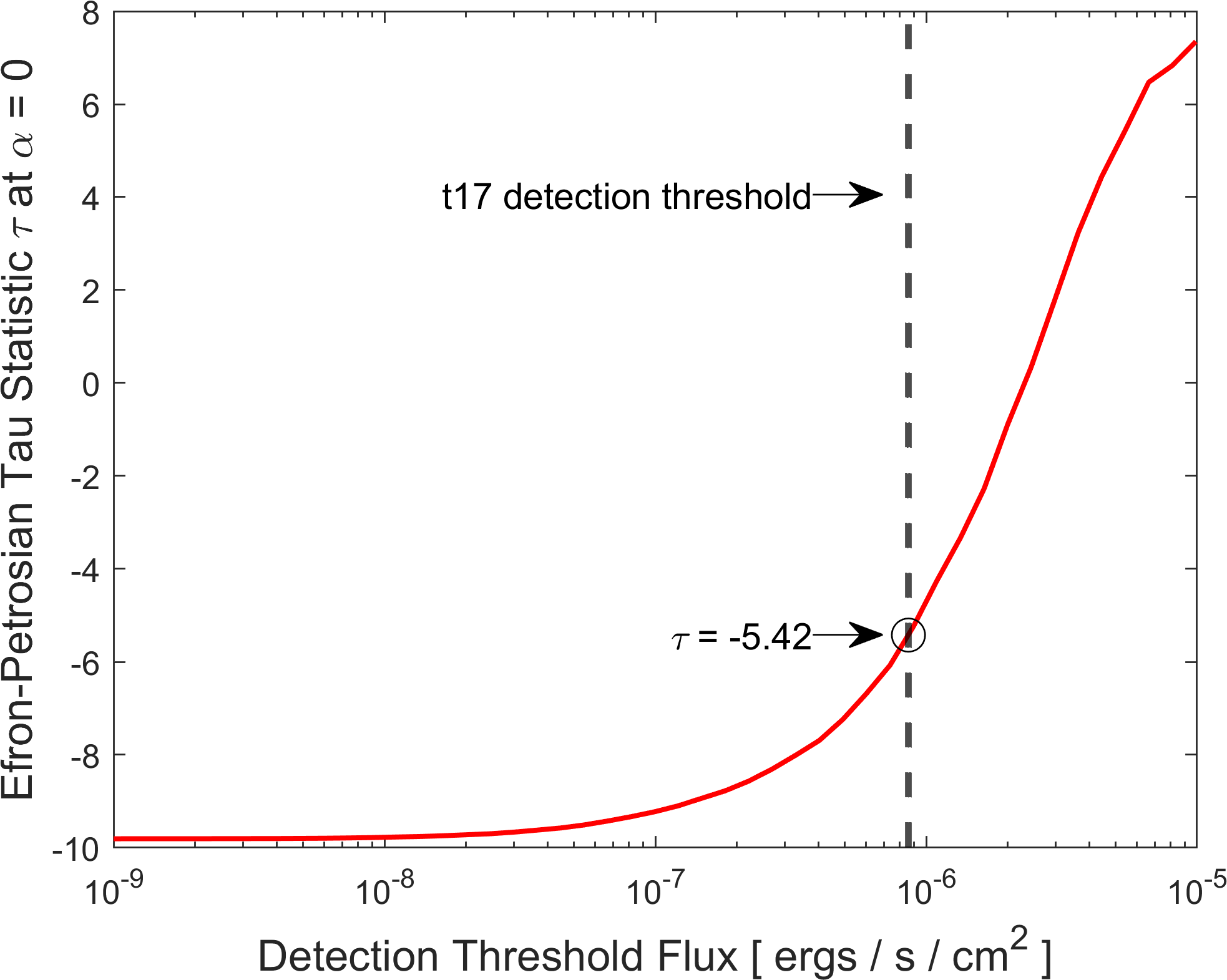} \label{fig:t17figsf}} \\
                \end{tabular}
            }
            \caption{
            Plotting of the 137 GRBs from \citetalias{tsvetkova2017konus}. Plot (a) shows redshift vs. isotropic luminosity. The black line represents the observational limit of \emph{Swift}, which has been deduced from \citetalias{tsvetkova2017konus} to be $8.6 \times 10^{-7}~[~erg~cm^{-2}~s^{-1}~]$. The purple line represents the linear regression through the data set whose slope is $\alpha=1.70$. Plot (b) is the observer frame visualization of the \citetalias{tsvetkova2017konus} data set, where the dashed line is the observational limit. Plot (c) is a histogram of flux, where the dashed line is the observational limit. Plot (d) shows redshift vs. $L_{0}=L(z)/(1+z)^{1.70}$, the redshift-independent luminosity. Plot (e) shows a range of possible threshold limits vs. $\alpha$ values at $\tau = 0$. The intersection of this line with \citetalias{tsvetkova2017konus}'s supposed threshold limit is our value for their $\alpha$. Plot (f) shows a range of possible threshold limits vs $\tau$ values at $\alpha = 0$. Assuming the detection threshold inferred from the analysis of \citetalias{tsvetkova2017konus}, a redshift-independent luminosity distribution is rejected at $5.4\sigma$. However, choosing a more conservative detection threshold at ~ $2\times10^{-6}~[~erg~cm^{-2}~s^{-1}~]$ yields no evidence for luminosity-redshift evolution.
            \label{fig:t17figs}
            }
        \end{figure*}

        In \citetalias{tsvetkova2017konus}, the authors explore a data set of GRBs detected in the triggered mode of the Konus-\emph{Wind} experiment. Beginning with 150 mixed-type GRBs, they prune the data set down to 137 by removing the Type I (short) GRBs as well as GRB 081203A. It is not explained why GRB 081203A is excluded.
        \newpar

        Similar to \citetalias{yu2015unexpectedly} and \citetalias{pescalli2016rate}, they employ the Efron-Petrosian $\tau$ statistic to find the luminosity evolution, assuming a functional form as seen in Eq. \eqref{eq:lumz}. They find individual truncation limits to the $(1+z)-\liso$ plane calculated for each burst separately, yielding an $\alpha$ value of $\alpha = 1.7$. They note that similar results were obtained by using the "monolithic" truncation limit of

        \begin{equation}
            \label{eq:fminT17}
            F_{min} = 2 \times 10^{-6}~[~erg~cm^{-2}~s^{-1}~] ~.
        \end{equation}

        \noindent Since both methods yield the same $\alpha$ value, we will use the single-valued limit, as we have in our previous analyses. When we do so, we obtain a value of $\alpha = 0.29$, which is significantly different from the reported value of 1.7. This should not be the case, so we turn to another method for obtaining $\alpha$ by visually matching the threshold cut from their Figure 8. Doing so requires a threshold of $1.1 \times 10^{-6}~[~erg~cm^{-2}~s^{-1}~]$ and yields $\alpha = 1.36.$ Still not the reported $\alpha$ value, we search for the threshold limit of $8.6 \times 10^{-7}~[~erg~cm^{-2}~s^{-1}~]$ which correctly yields $\alpha = 1.7$. This can be seen in Figures \ref{fig:t17figsa} - \ref{fig:t17figsc}.
        \newpar

        As can be seen in Figures \ref{fig:t17figsb} and \ref{fig:t17figsc}, the analysis of \citetalias{tsvetkova2017konus} also appears to suffer from an underestimation of the detector threshold limit of \emph{Swift}. Again, we expect the limit to be closer to the central peak of the histogram, soft-truncating the data set. Otherwise, such a sharp drop in the count of LGRB events before reaching the detection threshold would have truly fundamental and revolutionary implications about the cosmic rates of LGRBs.
        \newpar

        We note that \citetalias{tsvetkova2017konus}'s underestimation of the detector threshold limit does not appear to be as severe as \citetalias{yu2015unexpectedly} or \citetalias{pescalli2016rate}. Once the luminosity evolution has been removed, the detector threshold cut in the $(1+z)-L_{0}$ plane does not become flat at high redshift, as can be seen in Figure \ref{fig:t17figsd}.
        \newpar

        Also of note is the disparity in the significance of rejecting no luminosity-redshift evolution between our results and those of \citetalias{tsvetkova2017konus}. Figure \ref{fig:t17figsf} shows our inferred significance ($\tau\sim5.4\sigma$), which is hard to reconcile with the inferred significance $\tau = 1.2\sigma$ in \citetalias{tsvetkova2017konus}.

    \subsection{\citet{lloyd2019cosmological} (\citetalias{lloyd2019cosmological})}
    \label{sec:analysis:L19}

        \begin{figure*}
            \centering
            \makebox[\textwidth]
            {
                \begin{tabular}{ccc}
                    \subfloat[]{\includegraphics[width=0.31\textwidth]{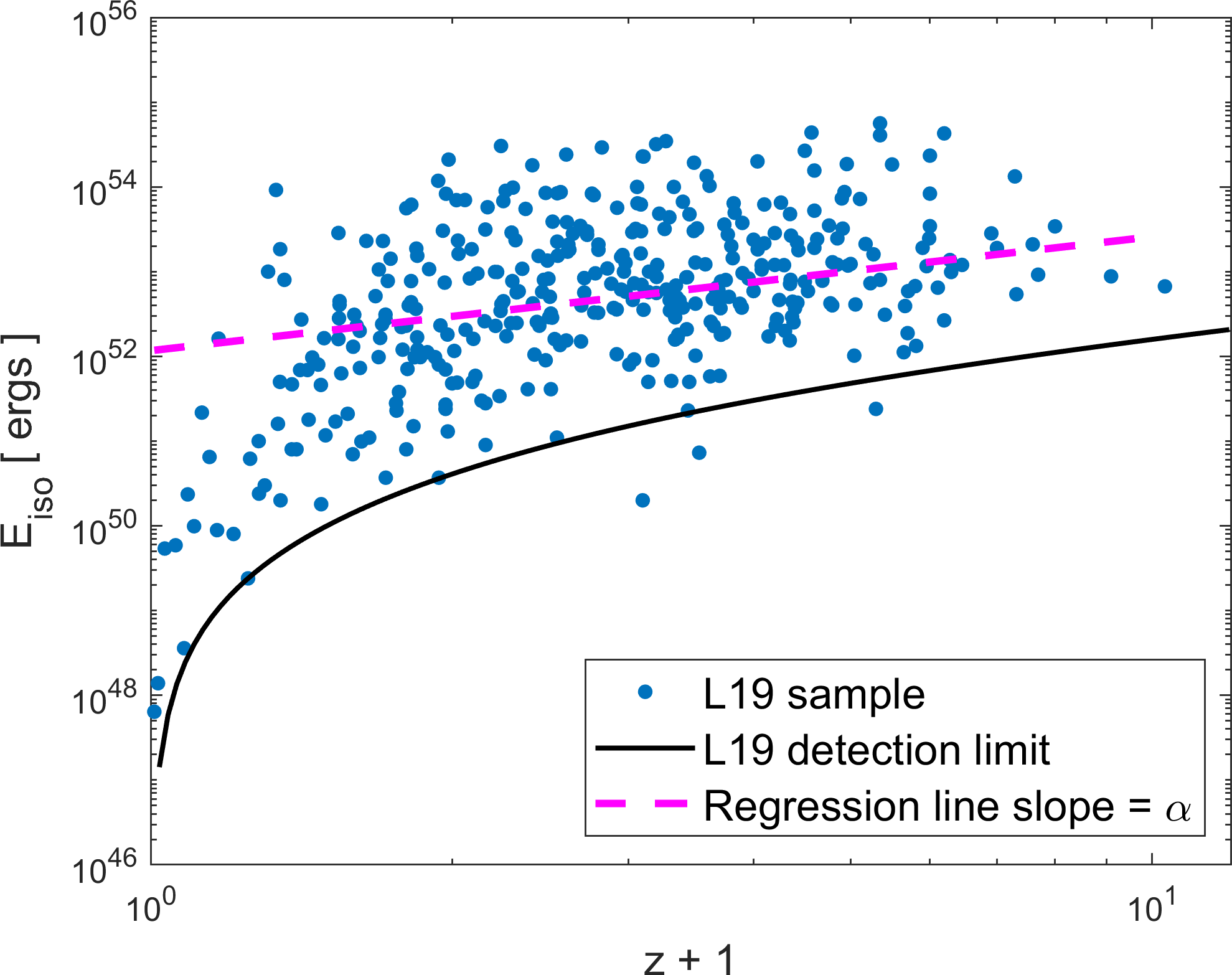} \label{fig:L19figsa}} &
                    \subfloat[]{\includegraphics[width=0.31\textwidth]{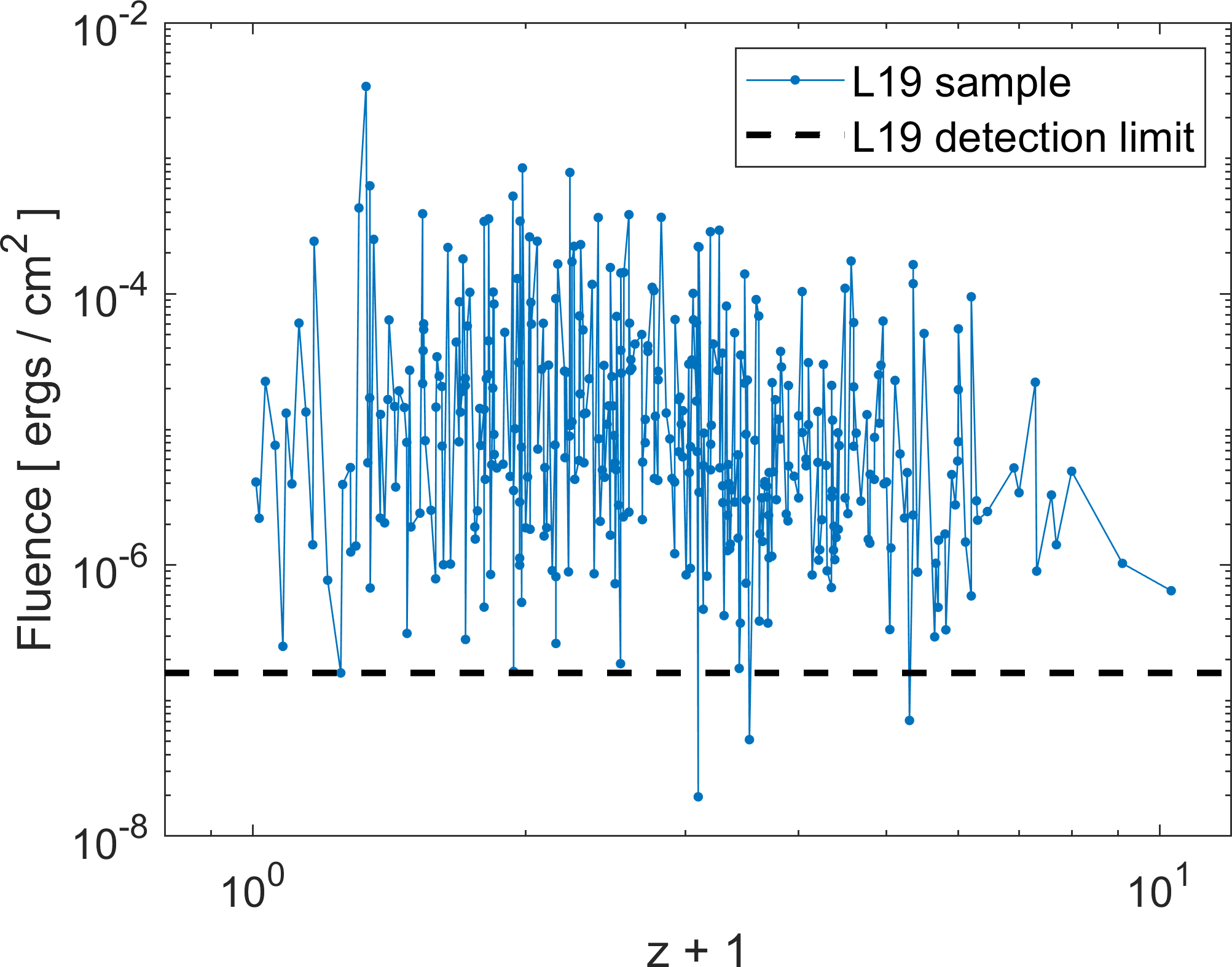} \label{fig:L19figsb}} &
                    \subfloat[]{\includegraphics[width=0.31\textwidth]{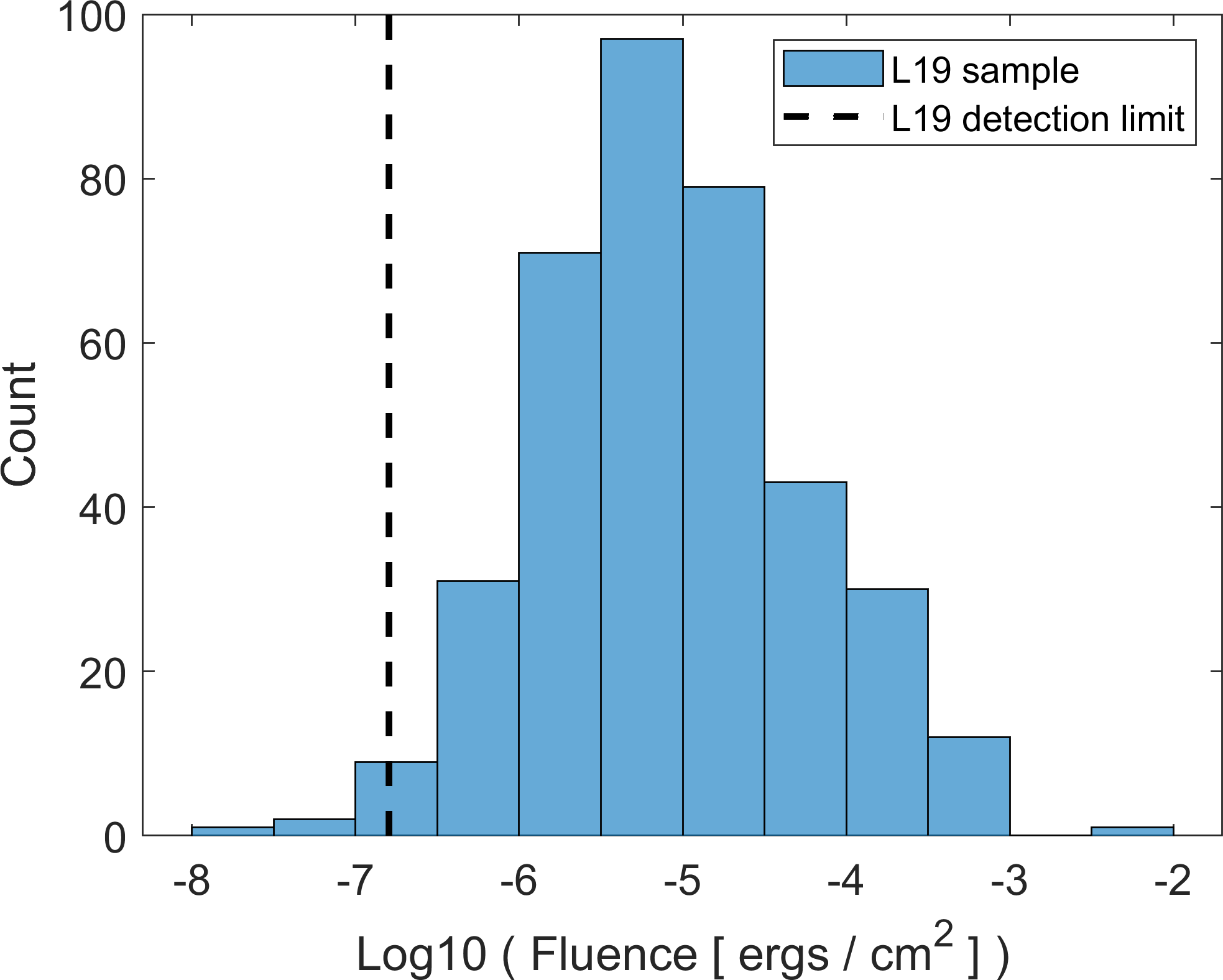} \label{fig:L19figsc}} \\
                    \subfloat[]{\includegraphics[width=0.31\textwidth]{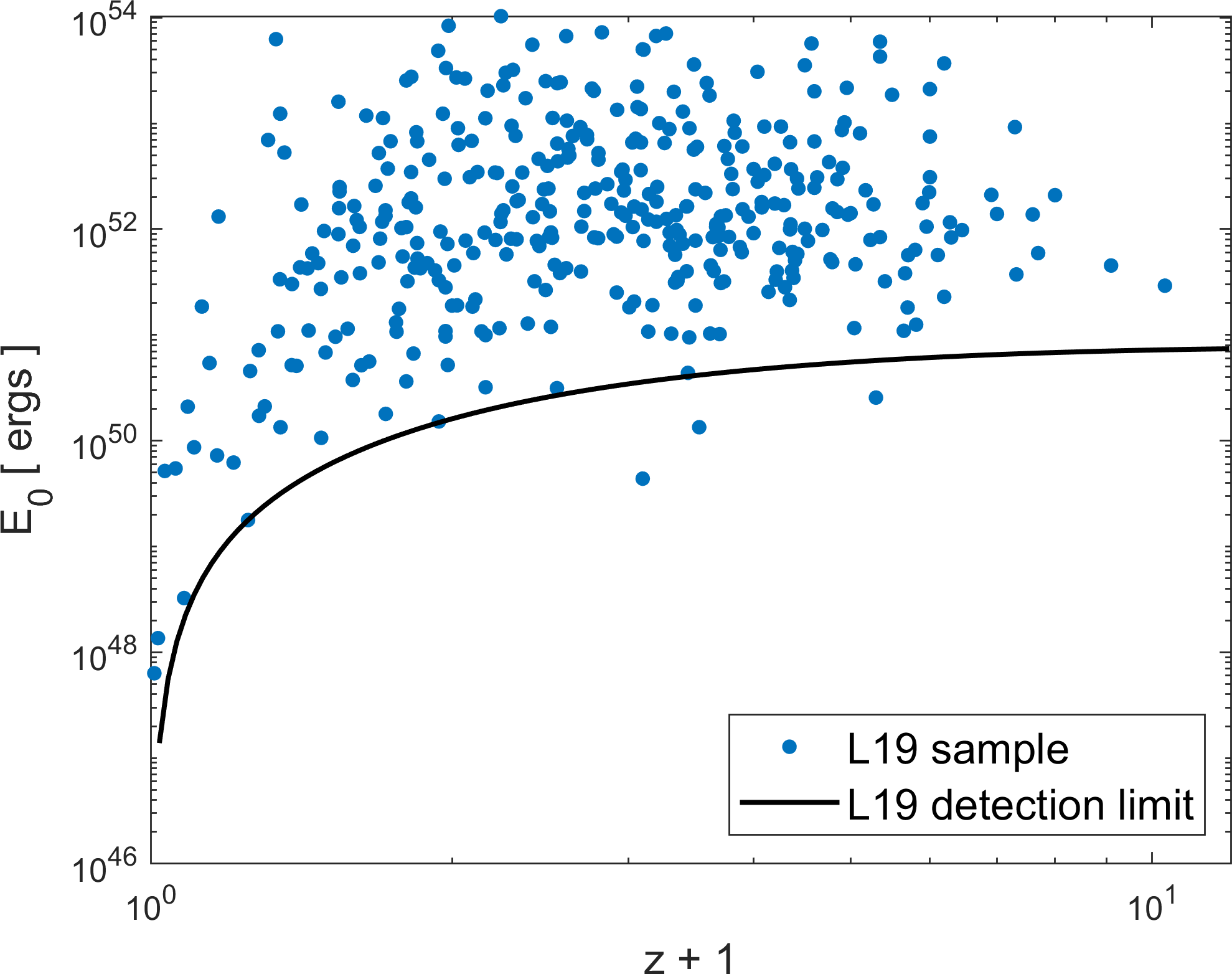} \label{fig:L19figsd}} &
                    \subfloat[]{\includegraphics[width=0.31\textwidth]{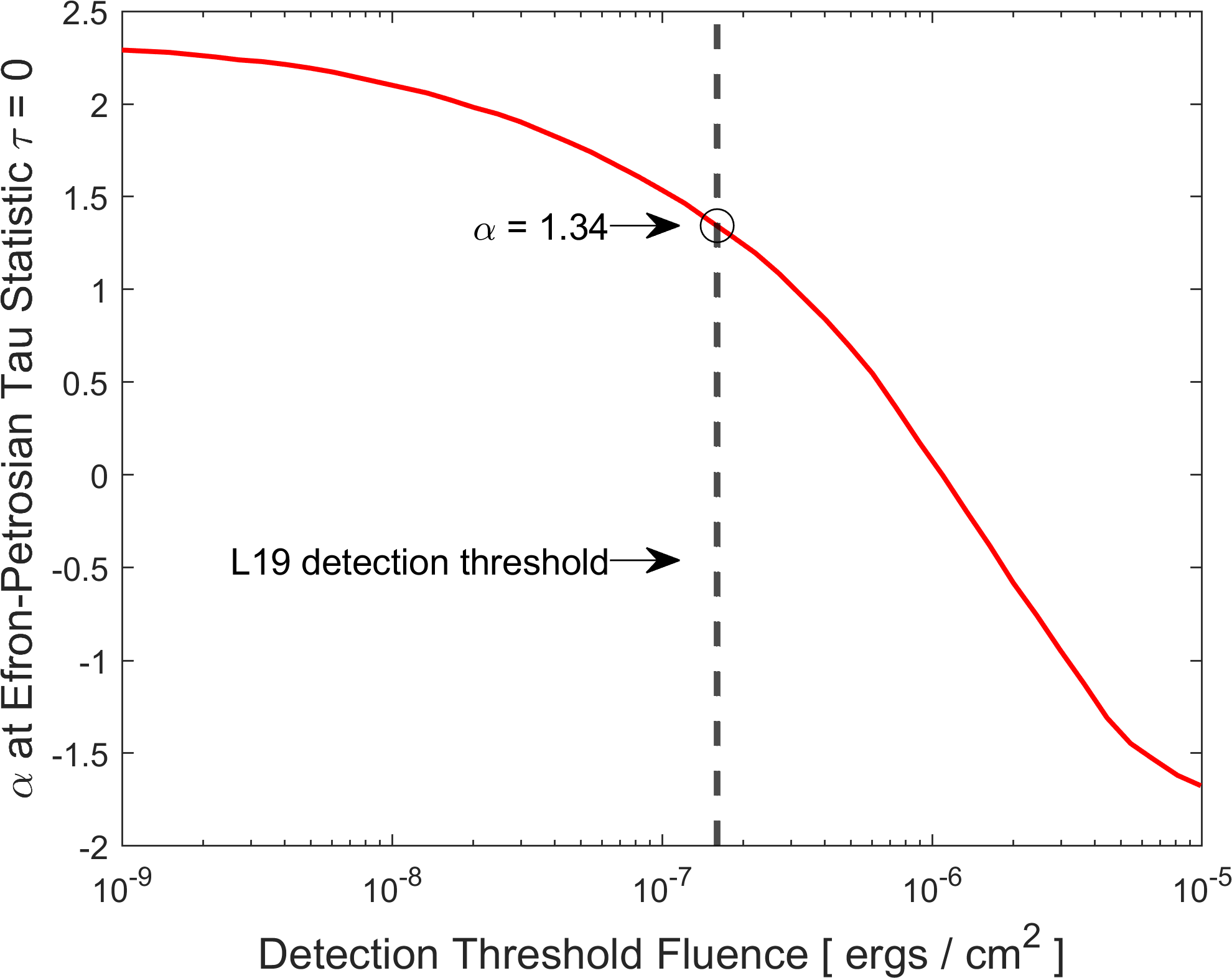} \label{fig:L19figse}} &
                    \subfloat[]{\includegraphics[width=0.31\textwidth]{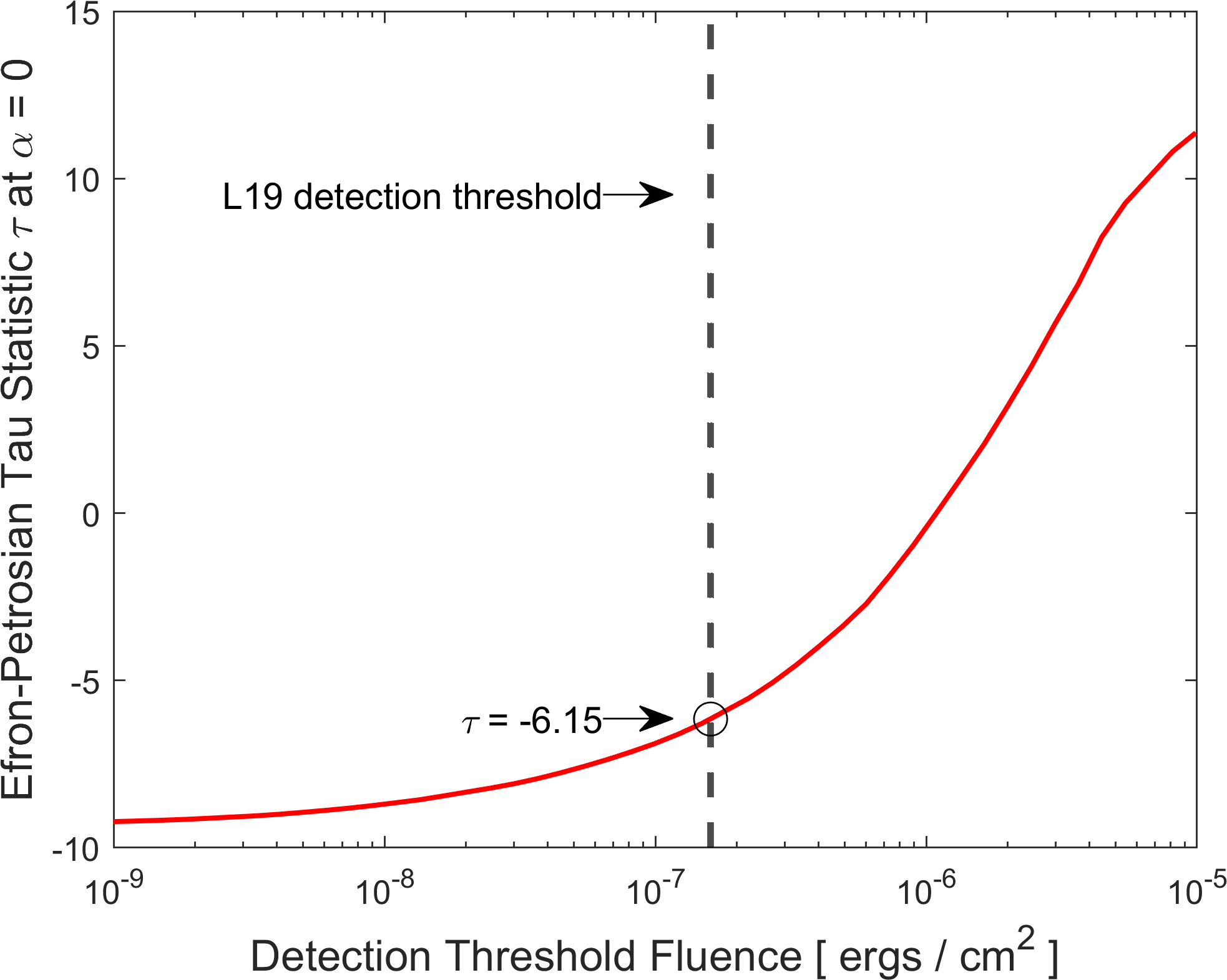} \label{fig:L19figsf}} \\
                \end{tabular}
            }
            \caption{
            Plotting of the 376 GRBs from \citetalias{lloyd2019cosmological}. Plot (a) shows redshift vs. isotropic emitted energy. The black line represents the observational limit of \emph{Swift}, which has been deduced from \citetalias{lloyd2019cosmological} to be $1.6 \times 10^{-7}~[~erg~cm^{-2}~]$. The purple line represents the linear regression through the data set whose slope is $\alpha=1.34$. Plot (b) is the observer frame visualization of the \citetalias{lloyd2019cosmological} data set, where the dashed line is the observational limit. Plot (c) is a histogram of fluence, where the dashed line is the observational limit. Plot (d) shows redshift vs. $E_{0}=E(z)/(1+z)^{1.34}$, the redshift-independent effective energy. Plot (e) shows a range of possible threshold limits vs. $\alpha$ values at $\tau = 0$. The intersection of this line with \citetalias{lloyd2019cosmological}'s supposed threshold limit is our value for their $\alpha$. Plot (f) shows a range of possible threshold limits vs $\tau$ values at $\alpha = 0$. Assuming the detection threshold inferred from the analysis of \citetalias{lloyd2019cosmological}, a redshift-independent luminosity distribution is rejected at $6.15\sigma$. However, choosing a more conservative detection threshold at ~ $\times10^{-6}~[~erg~cm^{-2}~]$ yields no evidence for luminosity-redshift evolution.
            \label{fig:L19figs}
            }
        \end{figure*}

        In \citetalias{lloyd2019cosmological}, the authors use a data set taken from \citet{wang2020comprehensive} that consists of all publicly available observations of 6289 GRBs, from 1991 to 2016. They isolate those events with a measured redshift and observed duration of $T_{90} > 2s$ (LGRBs), which can yield an estimate of the total isotropic gamma-ray emission, $E_{iso}$. This leads to the selection of 376 LGRB events by \citetalias{lloyd2019cosmological}, based upon which they proceed to construct the $\tau$ statistic, in similar fashion to the previous studies, to find the redshift evolution parameter, $\alpha$. \citetalias{lloyd2019cosmological} choose a functional form of $E_{iso} = E_{0}(1+z)^{\alpha}$, and find that $\alpha = 2.3 \pm 0.5$. \citetalias{lloyd2019cosmological} report a value of,

        \begin{equation}
            \label{eq:fminT17}
            F_{min} = 2 \times 10^{-6}~[~erg~cm^{-2}~] ~.
        \end{equation}

        \noindent used in their study. However, when we use this limit in our reanalysis, the threshold line cuts through the majority of the data set, and yields a value of $\alpha = -0.58$. Our inferred value for $\alpha$ is completely as odds with the value reported by \citetalias{lloyd2019cosmological}. In order to obtain their value of $\alpha = 2.3$, we have to use a threshold limit of $7.0 \times 10^{-10}~[~erg~cm^{-2}~]$. This inferred detection threshold cut is almost two orders of magnitude below the data set.
        \newpar

        To resolve the disagreement between our inferred value of detection threshold used by \citetalias{lloyd2019cosmological} and the reported value in their study, we instead settle on visually matching the threshold cut in Figure 1 of \citetalias{lloyd2019cosmological} to obtain a limit of $1.6 \times 10^{-7}~[~erg~cm^{-2}~]$. This yields Figure \ref{fig:L19figsa} which looks remarkably similar to the corresponding Figure 1 of \citetalias{lloyd2019cosmological}. Assuming this detection threshold, we obtain a value of $\alpha = 1.34$ using the Efron-Petrosian $\tau$ statistic.
        \newpar

        If our inferred the detection threshold is indeed the value used by \citetalias{lloyd2019cosmological} in their study, then Figures \ref{fig:L19figsb} and \ref{fig:L19figsc}, lead us to conclude that the detection threshold has been likely severely underestimated in the study of \citetalias{lloyd2019cosmological}. The effective threshold represented by the dashed line in the histogram of Figure \ref{fig:L19figsc} is well to the left of the peak of the distribution of LGRB fluences.
        \newpar

        Our conclusion in the above is further confirmed by Figure \ref{fig:L19figsd}, where we plot the redshift-corrected isotropic effective energy vs. redshift. The solid black line in this plot represents the redshift-corrected detection threshold and almost resembles a flat line at high redshifts. This is another indication that the inferred relationship between (1+z) and $E_{iso}$ is likely heavily influenced by the improperly-modeled detection threshold. In this study, however, the effective threshold cut represent the combined effects of the detection thresholds of multiple satellites due to the heterogenous collection of data from multiple independent GRB catalogs. All of these raise the possibility that \citetalias{lloyd2019cosmological}'s single-valued threshold is likely a severe oversimplification of the complex merger of individual satellite thresholds, while none of the individual satellite thresholds might be well represented by a single-valued hard cutoff.
        \newpar

        Figure \ref{fig:L19figse} shows a range of possible threshold limits  vs. $\alpha$ values at $\tau = 0$. One can see the red line approaches \citetalias{lloyd2019cosmological}'s value of $\alpha = 2.3$ on the far left edge of the graph. Figure \ref{fig:L19figsf} indicates that $E_{iso}$ and $(1+z)$ are correlated at $6.15\sigma$ significance assuming our inferred threshold has been used by \citetalias{lloyd2019cosmological}. Although this significance is not the same as the value reported by \citetalias{lloyd2019cosmological}, $5.6\sigma$, it is consistent. Note, however, that this inferred detection threshold still severely underestimates the actual combined effects of multiple detection thresholds on the heterogenous data set of \citetalias{lloyd2019cosmological}.

\section{Monte Carlo Simulations of Luminosity-Redshift Evolution}
\label{sec:simulation}

    \begin{figure*}
        \centering
        \makebox[\textwidth]
        {
            \begin{tabular}{ccc}
            \subfloat[]{\includegraphics[width=0.49\textwidth]{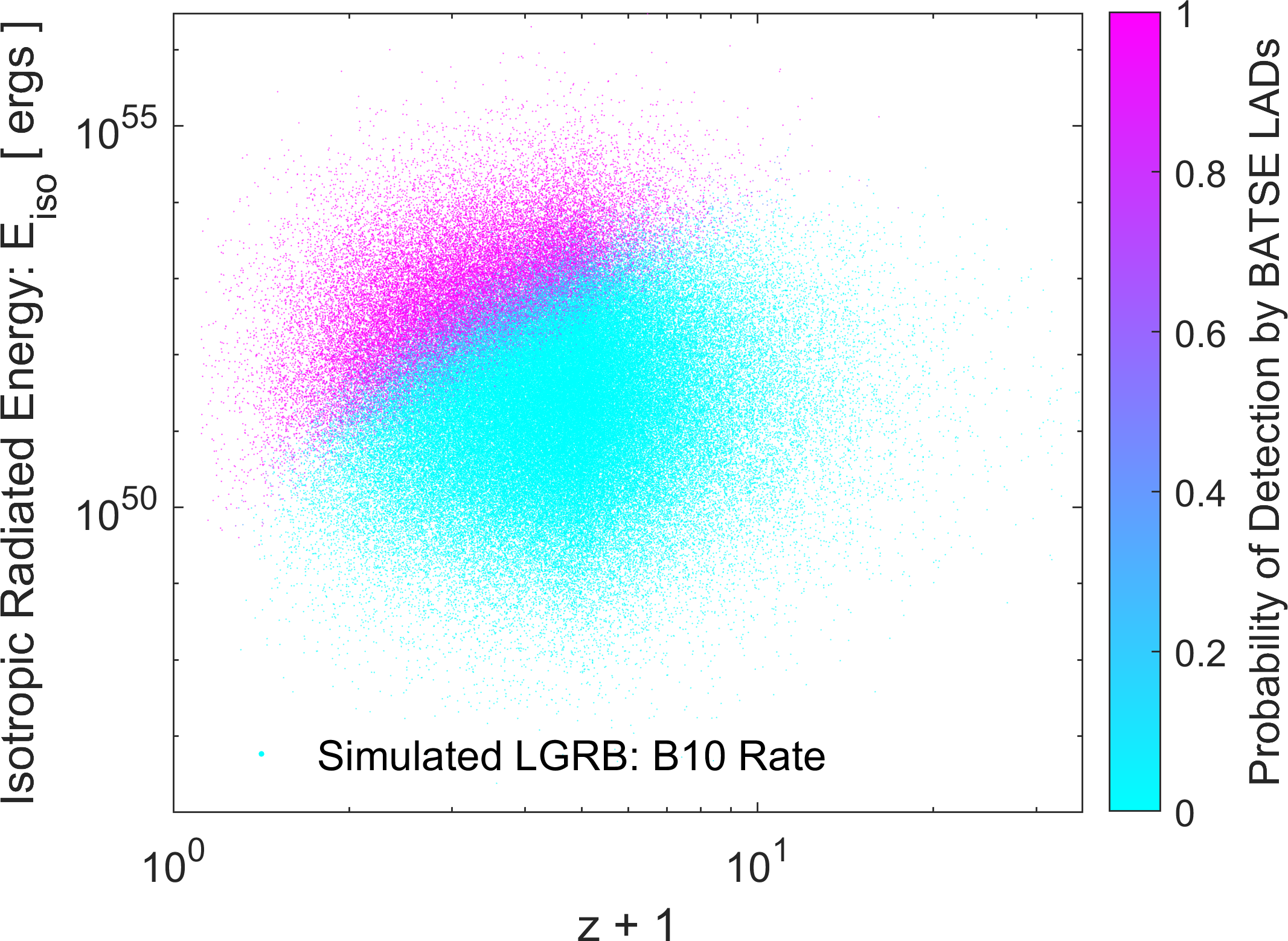} \label{fig:SynEisoRedshift}} &
            \subfloat[]{\includegraphics[width=0.49\textwidth]{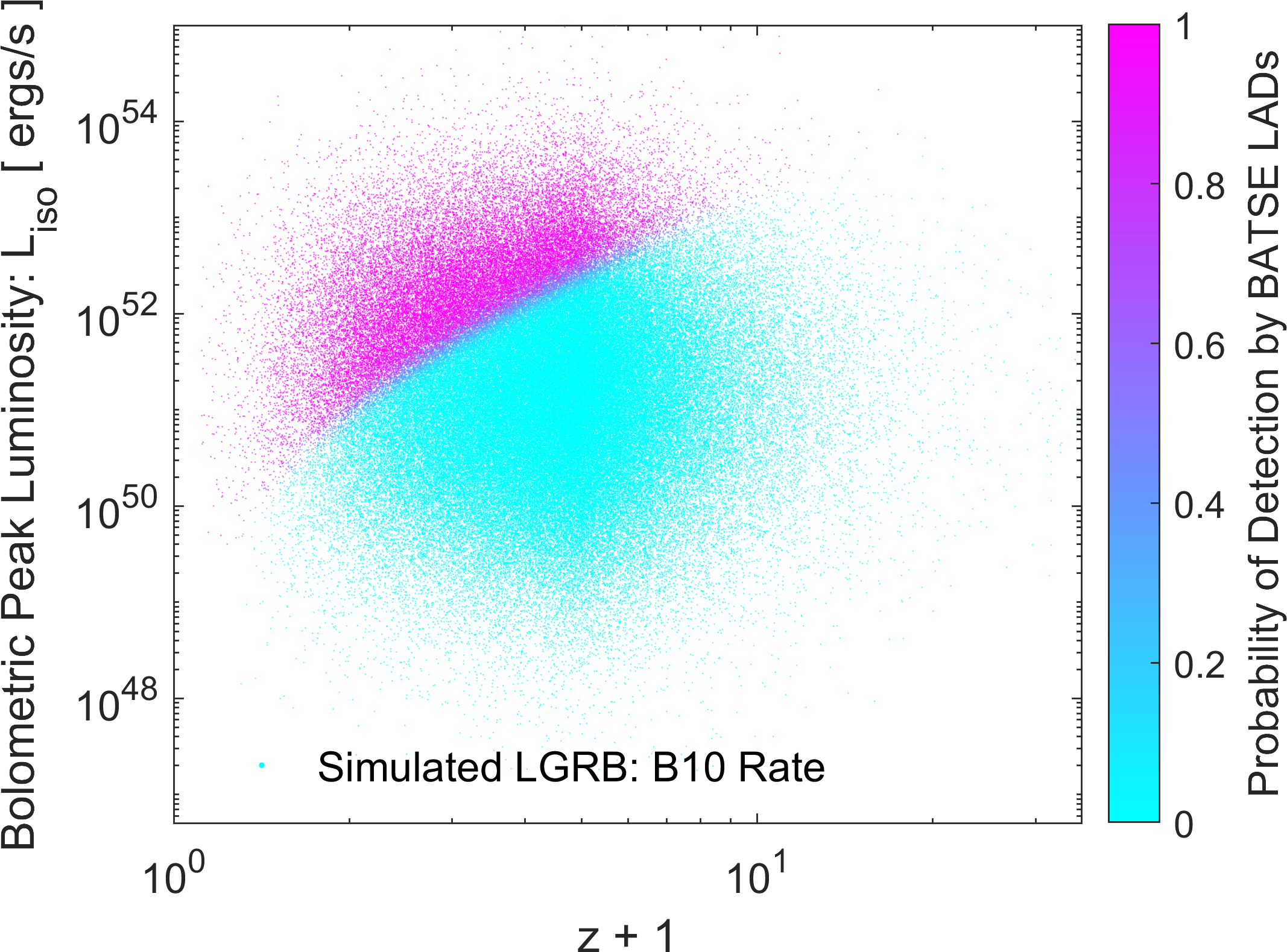} \label{fig:SynLisoRedshift}} \\
            \end{tabular}
        }
        \caption{
        An illustration of the Monte Carlo universe of LGRBs constructed in \S\ref{sec:simulation:MonteCarloGRBWorldModel}. Each point in this plot represents one synthetic LGRB. The magenta color represents a high probability of detection while the cyan represents a low probability of detection.
        \label{fig:SynFullData}
        }
    \end{figure*}

    \begin{figure*}
        \centering
        \makebox[\textwidth]
        {
            \begin{tabular}{cccc}
            \subfloat[]{\includegraphics[width=0.49\textwidth]{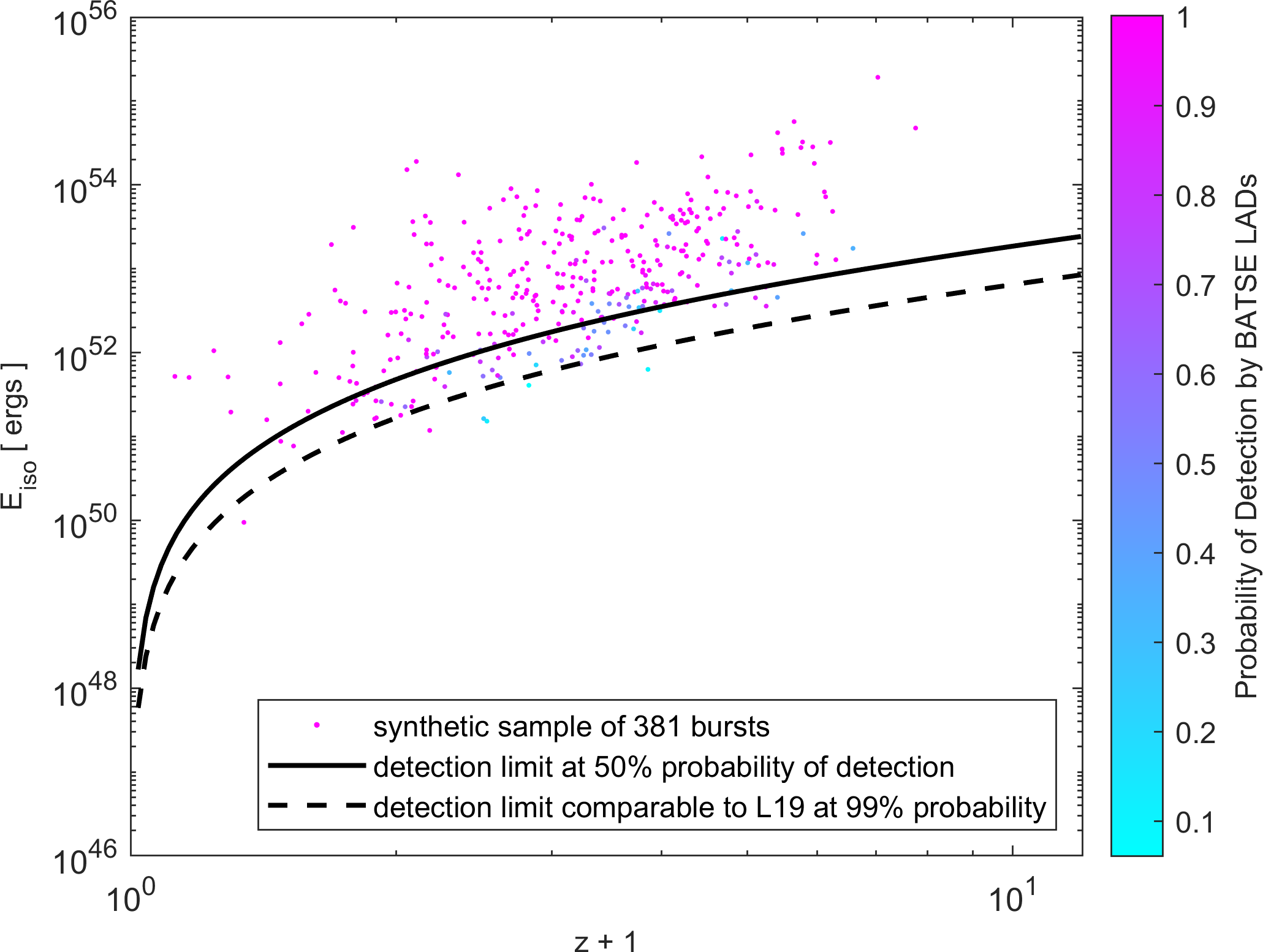} \label{fig:synSamFluenceZoneEiso}} &
            \subfloat[]{\includegraphics[width=0.49\textwidth]{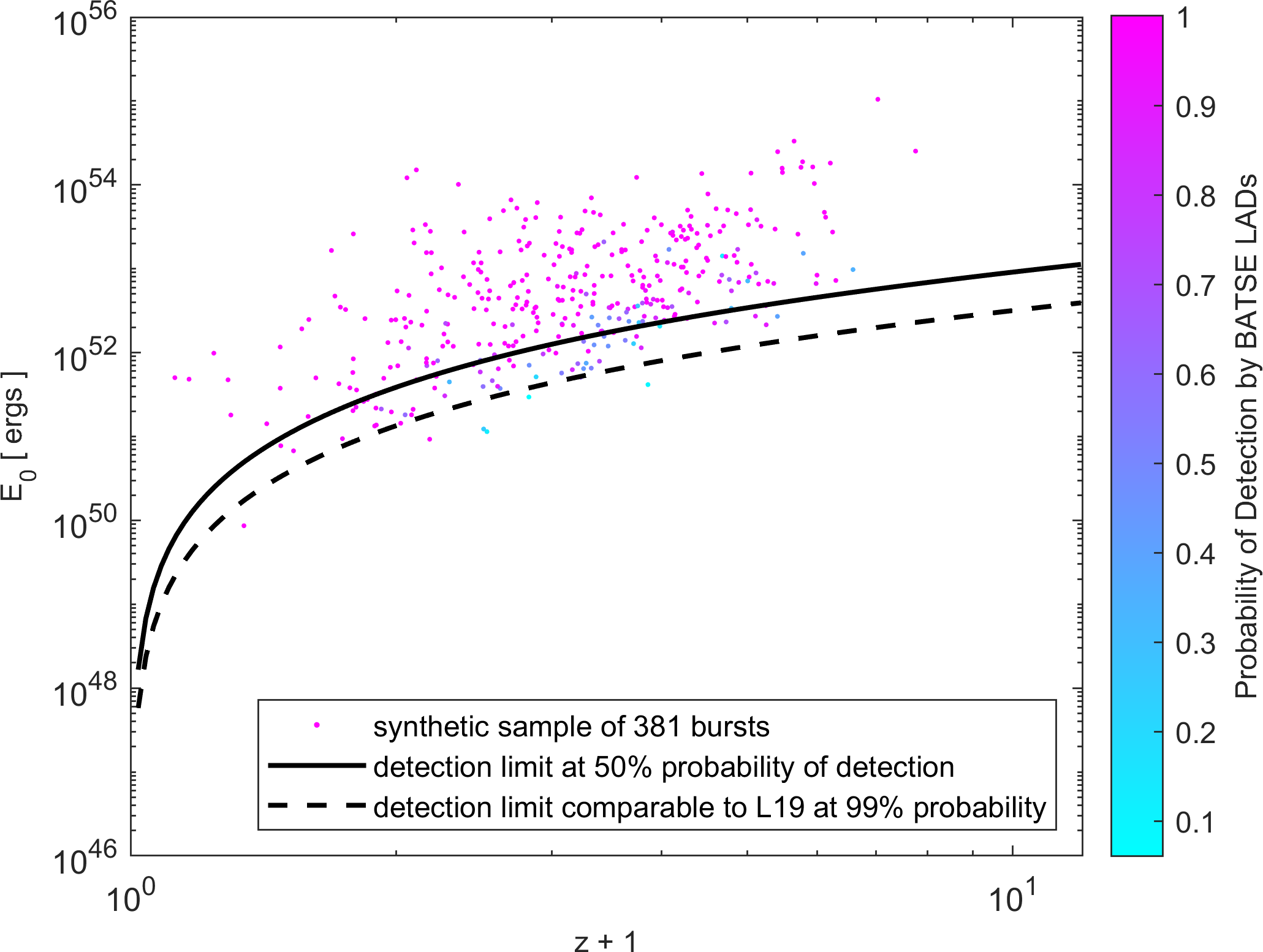} \label{fig:synSamFluenceZoneEisoCorrected}} \\
            \subfloat[]{\includegraphics[width=0.49\textwidth]{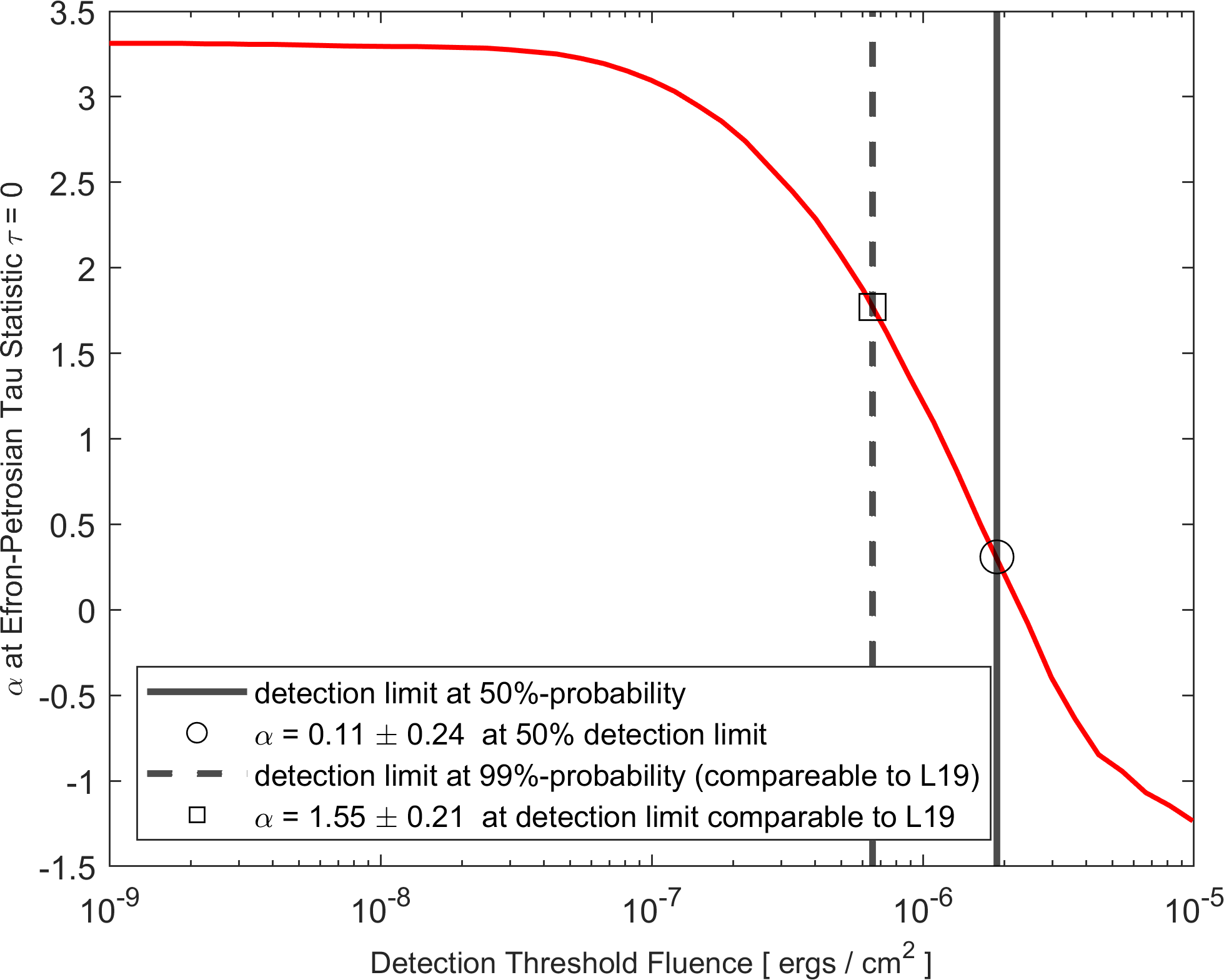} \label{fig:synSamFluenceThreshAlpha}} &
            \subfloat[]{\includegraphics[width=0.49\textwidth]{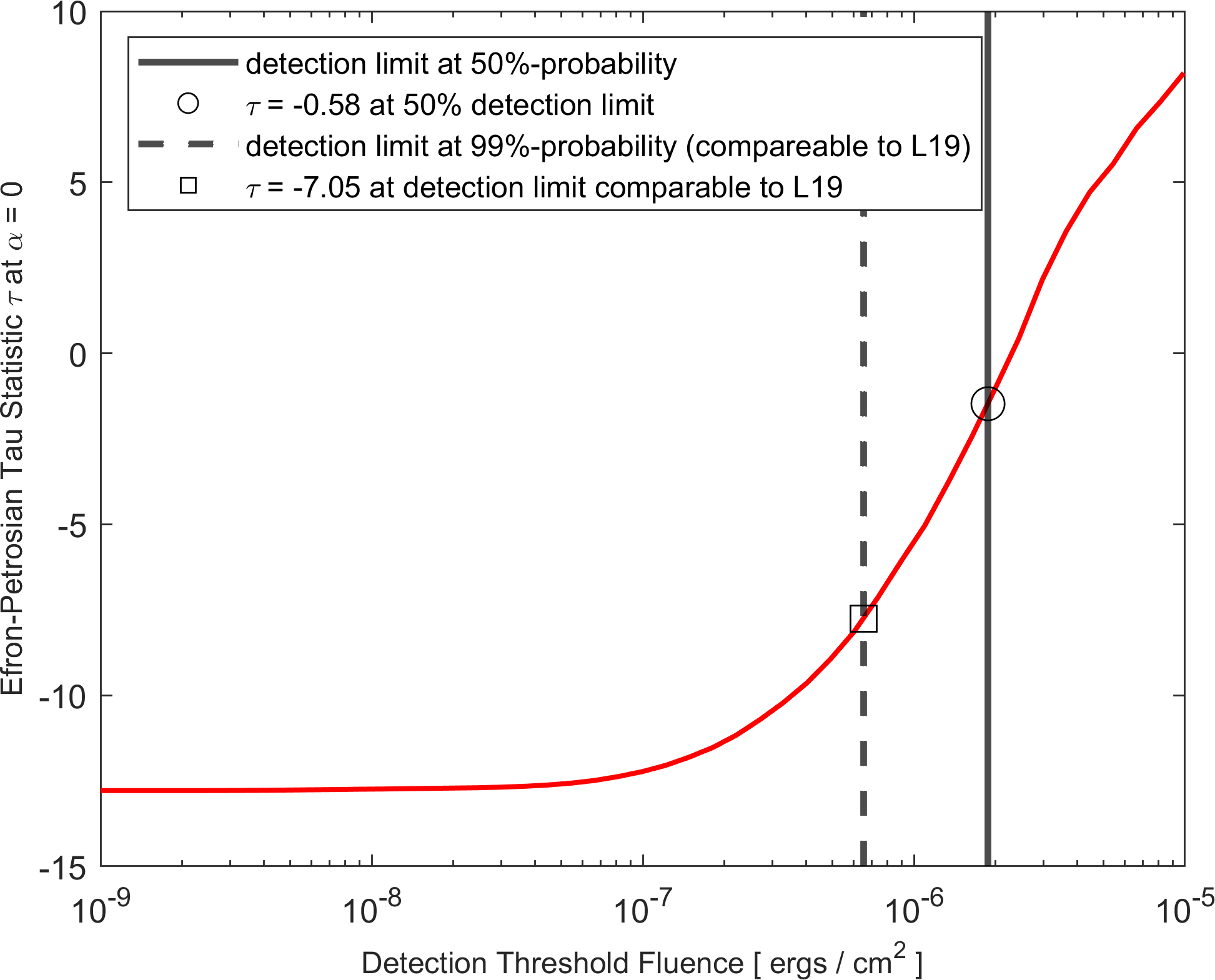} \label{fig:synSamFluenceThreshTau}} \\
            \end{tabular}
        }
        \caption{
        An illustration of the effects of detection threshold on the outcome of the Efron-Petrosian test statistic. Plot {\bf (a)} Shows the synthetic data set used for our study of $(\eiso,z+1)$ correlation. The solid black line represents the detector threshold at 50\% while the dashed black line represents a detector threshold comparable to that of \citetalias{lloyd2019cosmological} at 99\% probability of detection. The color bar represents the probability of detection by the BATSE LADs where cyan and magenta represent 0\% and 100\% chances of detection, respectively. Plot {\bf (b)} shows the redshift-evolution corrected data set based off of the value of alpha calculated using the detector threshold at 50\% probability of detection. Plot {\bf (c)} shows the alpha value calculated corresponding to $\tau = 0$ with varying detection threshold limits. The solid black line represents the detector threshold at 50\%, the black circle is the average $\alpha$ value over 50 generated samples for the specific threshold used, while the dashed black line represents the detector threshold at 99\% probability of detection, comparable to those of previous studies, and the black square is the average $\alpha$ over 50 generated samples at $\tau = 0$. Plot {\bf (d)} displays the $\tau$ statistic at $\alpha = 0$. The black line represents the detector threshold at 50\% detection probability and the dashed black line represents the detection threshold at 99\% detection probability, comparable to those of previous studies. The circle and square in this figure are the average $\tau$ values over 50 generated samples.
        \label{fig:synSamEiso}
        }
    \end{figure*}

    \begin{figure*}
        \centering
        \makebox[\textwidth]
        {
            \begin{tabular}{cccc}
            \subfloat[]{\includegraphics[width=0.49\textwidth]{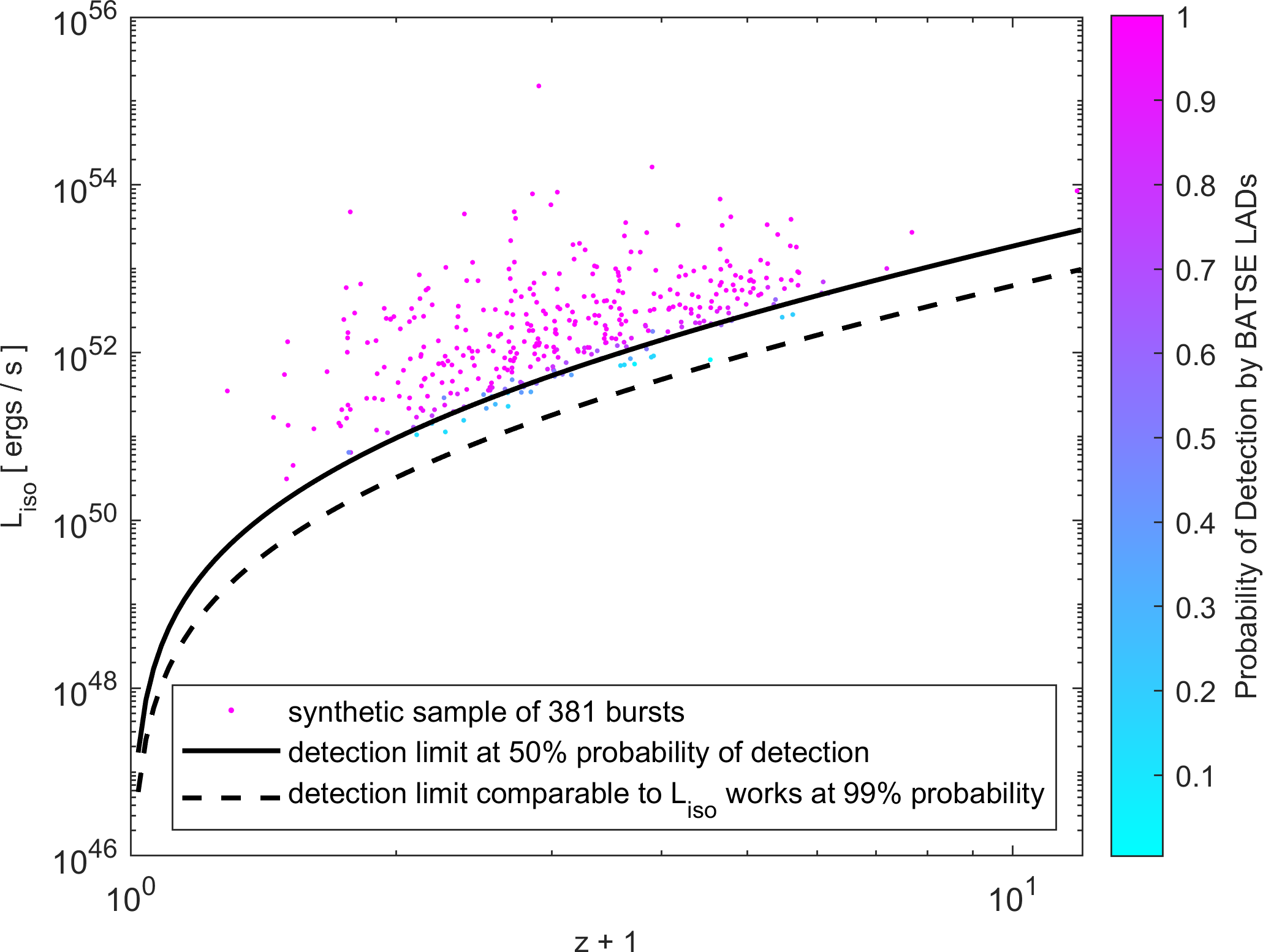} \label{fig:synSamFluxZoneLiso}} &
            \subfloat[]{\includegraphics[width=0.49\textwidth]{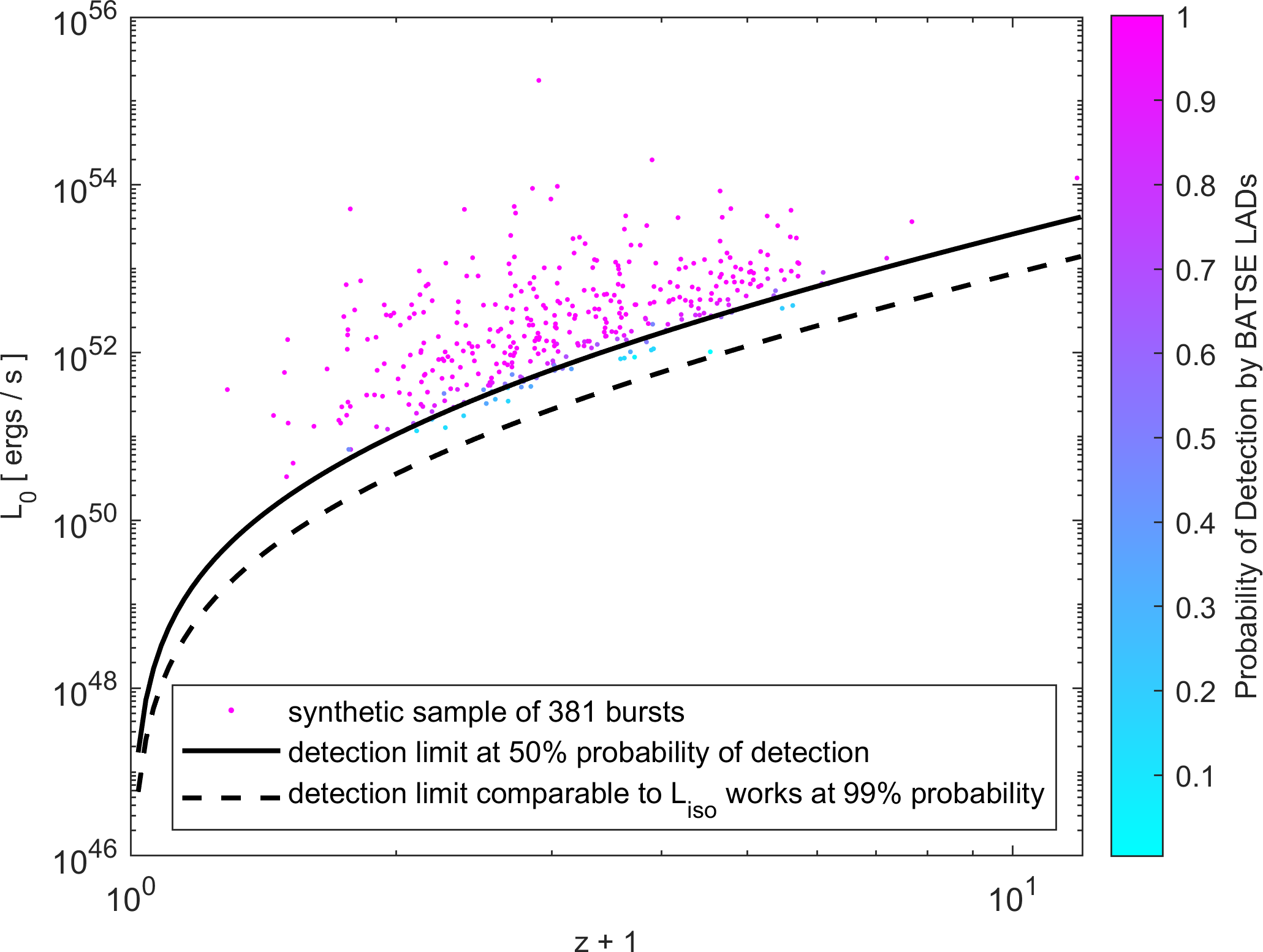} \label{fig:synSamFluxZoneLisoCorrected}} \\
            \subfloat[]{\includegraphics[width=0.49\textwidth]{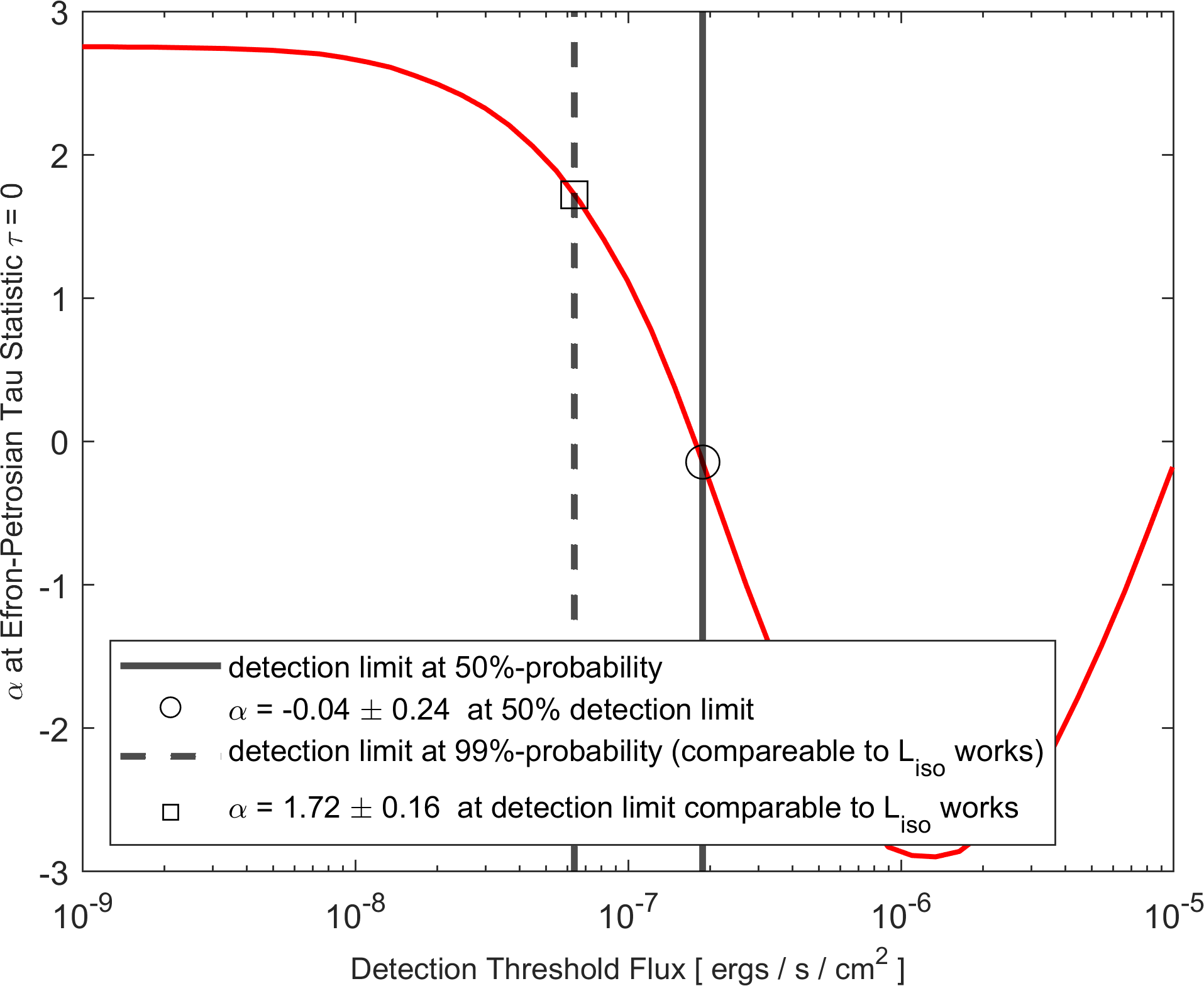} \label{fig:synSamFluxThreshAlpha}} &
            \subfloat[]{\includegraphics[width=0.49\textwidth]{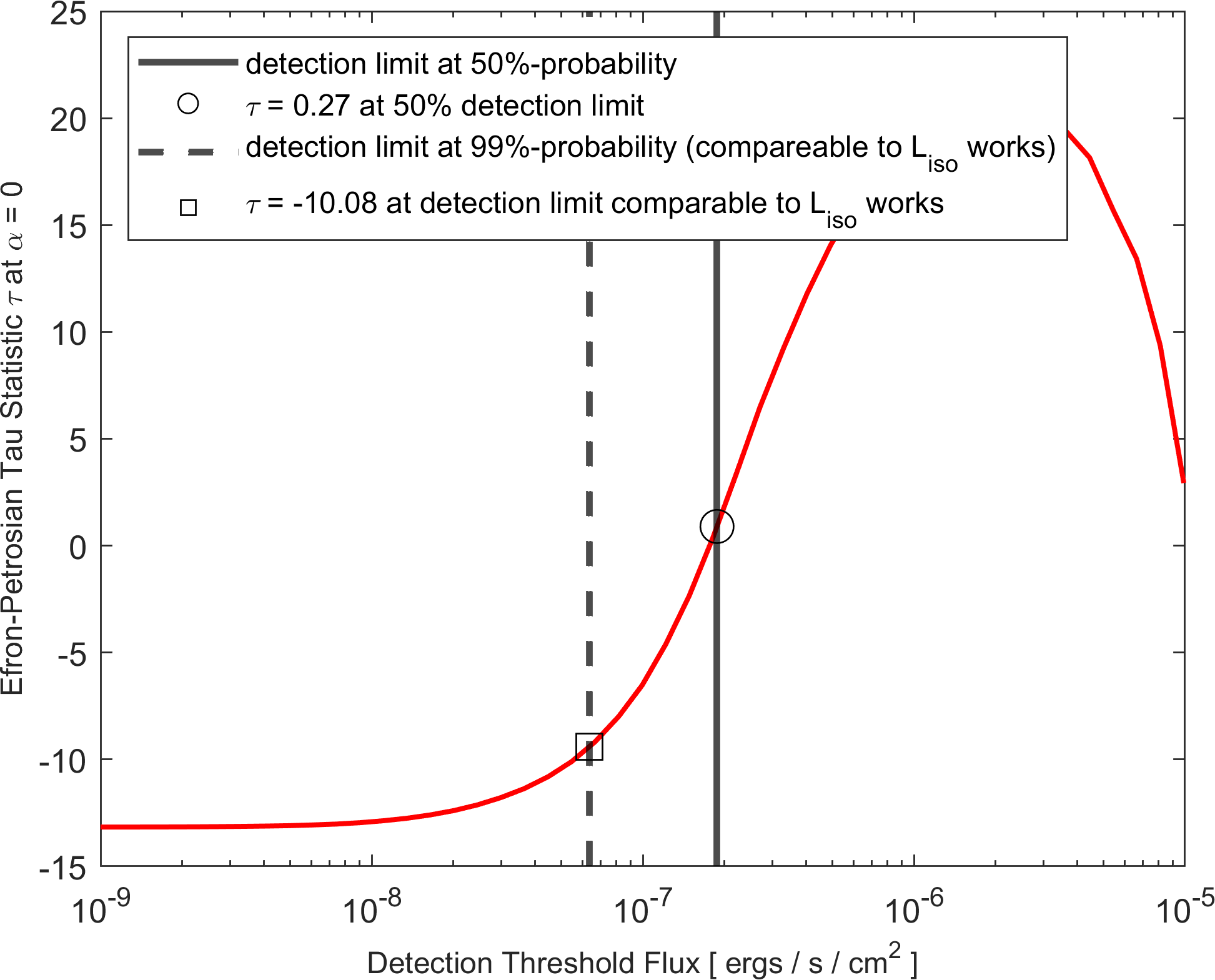} \label{fig:synSamFluxThreshTau}} \\
            \end{tabular}
        }
        \caption{
        An illustration of the effects of detection threshold on the outcome of the Efron-Petrosian test statistic. Plot {\bf (a)} Shows the synthetic data set used for our study of $(\liso,z+1)$ correlation. The solid black line represents the detector threshold at 50\% while the dashed black line represents a detector threshold comparable to that of \citetalias{lloyd2019cosmological} at 99\% probability of detection. The color bar represents the probability of detection by the BATSE LADs where cyan and magenta represent 0\% and 100\% chances of detection, respectively. Plot {\bf (b)} shows the redshift-evolution corrected data set based off of the value of alpha calculated using the detector threshold at 50\% probability of detection. Plot {\bf (c)} shows the alpha value calculated corresponding to $\tau = 0$ with varying detection threshold limits. The solid black line represents the detector threshold at 50\%, the black circle is the average $\alpha$ value over 50 generated samples for the specific threshold used, while the dashed black line represents the detector threshold at 99\% probability of detection, comparable to those of previous studies, and the black square is the average $\alpha$ over 50 generated samples at $\tau = 0$. Plot {\bf (d)} displays the $\tau$ statistic at $\alpha = 0$. The black line represents the detector threshold at 50\% detection probability and the dashed black line represents the detection threshold at 99\% detection probability, comparable to those of previous studies. The circle and square in this figure are the average $\tau$ values over 50 generated samples.
        \label{fig:synSamLiso}
        }
    \end{figure*}

    To further investigate the effects of detection threshold underestimation on the inferred luminosity-redshift evolution of LGRBs, we also create a Monte Carlo universe of LGRBs. The premise of our simulation is that the LGRB world is devoid of any luminosity-redshift or energetics-redshift correlations. Therefore, the application of the Efron-Petrosian statistic on any collection of events measured from such Monte Carlo universe of LGRBs, subjected to a given simulated detection threshold, should also accurately predict no energetics/luminosity-redshift correlations for the intrinsic underlying population of LGRBs in the Monte Carlo universe.

    \subsection{The LGRB World Model}
    \label{sec:simulation:GRBWorldModel}

        We follow the approach extensively discussed in \citet{shahmoradi2013multivariate, shahmoradi2013gamma, shahmoradi2015short,  shahmoradi2019catalog, 2019arXiv190306989S, 2020arXiv200601157O}, to construct our Monte Carlo Universe of LGRBs. Toward this, we assume that the intrinsic comoving rate density of LGRBs follows a piecewise power-law function of the form,

        \begin{equation}
            \label{eq:mz}
            \mz(z) \propto
            \begin{cases}
                (1+z)^{\gamma_0} & z<z_0 \\
                (1+z)^{\gamma_1} & z_0<z<z_1 \\
                (1+z)^{\gamma_2} & z>z_1 ~, \\
            \end{cases}
        \end{equation}

        \noindent whose parameters, 
        
        \begin{equation}
            \label{eq:mz}
            (z_0,z_1,\gamma_0,\gamma_1,\gamma_2) = (0.97,4.00,3.14,1.36,-2.92) ~,
        \end{equation}

        \noindent are adopted from \cite{butler2010cosmic}. For simplicity, but also as argued by \citet{shahmoradi2013multivariate, shahmoradi2013gamma, shahmoradi2015short,  shahmoradi2019catalog, 2019arXiv190306989S, 2020arXiv200601157O}, we consider a 4-dimensional Multivariate Lognormal Probability Density Function (PDF) for the joint distribution of the four main LGRB prompt gamma-ray emission characteristics: The total isotropic peak luminosity ($\liso$), the total gamma-ray emission ($\eiso$), the intrinsic spectral peak energy ($\epkz$), and the intrinsic duration ($\durz$).
        \newpar

        We use the BATSE catalog of 1366 LGRBs \citep{paciesas1999fourth, shahmoradi2010hardness, goldstein2013batse} to construct a Bayesian hierarchical model \citep{shahmoradi2017multilevel, 2017arXiv171110599S} of the cosmic distribution of LGRBs in the universe, subject to an accurate modeling of the the detection threshold of BATSE Large Area Detectors (LADs) and data uncertainties. Then, we use a variant of the adaptive Markov Chain Monte Carlo techniques to sample the resulting posterior distribution of the parameters of the hierarchical model \citep{shahmoradi2019paramonte, shahmoradi2020paradram, 2020arXiv200809589S, 2020arXiv201000724S, 2020arXiv200914229S, 2020arXiv200914229S}. Details of model construction and sampling are extensively discussed in the aforementioned papers \citep[e.g.,][]{shahmoradi2013multivariate, 2020arXiv200601157O}.

    \subsection{The Monte Carlo Universe of LGRBs}
    \label{sec:simulation:MonteCarloGRBWorldModel}

        Once the best-fit parameters of the LGRB world model are inferred, we create a Monte Carlo Universe of LGRBs by randomly and repeatedly generating LGRB events whose characteristics are distributed according to the LGRB World model constructed in \S\ref{sec:simulation:GRBWorldModel}. For each LGRB event synthesis, we use a set of model parameters randomly drawn from the posterior distribution of the LGRB world model parameters explored in \S\ref{sec:simulation:GRBWorldModel}. Then, each LGRB passes through the simulated LGRB detection process of the BATSE LADs.
        \newpar

        An illustration of the resulting Monte Carlo Universe of LGRBs is provided in Figure \ref{fig:SynFullData}. The two plots represent the joint distributions of $\eiso$/$\liso$ and redshift. Clearly, the BATSE LADs create a rather sharp cut on the synthesized $z-\liso$ sample of LGRBs compared to the distribution of $z-\eiso$, which exhibits much fuzzier detection threshold effects. This is expected and reassuring, since the BATSE LADs primarily triggered on the peak photon flux at different timescales.
        \newpar

        We note that the specific shape of the energetics or redshift distribution of LGRBs or the specific detection mechanism of LGRBs in our Monte Carlo simulations has no relevance or effects on our assessment of the utility and accuracy of the Efron-Petrosian statistic. All that is important here, is the lack of any a priori correlations between the energetics and the redshifts of LGRBs in our Monte Carlo simulations.
        \newpar

        Using our Monte Carlo universe of LGRBs, we generate a random sample of 380 BATSE-detectable LGRBs. This sample size is comparable to the size of the observational data sets collected and analyzed in previous studies. We flag an LGRB as detectable by generating a uniform-random number between 0 and 1 and comparing it to the probability of detection of the LGRB. If the probability of detection is higher than the randomly generated number, we include the event in the sample of detected LGRBs for our analysis.

    \subsection{Analysis of Synthetic Monte Carlo Data}
    \label{sec:simuilation:Analysis}

        We start with the synthetic $\eiso-z$ sample shown in figure \ref{fig:synSamFluenceZoneEiso} where the black line represents the BATSE detector threshold at 50\% detection probability and the color on each point represents the probability of that burst being detected by BATSE. The corresponding lower limit on the fluence at $50\%$ chance of detection is

        \begin{equation}
            \label{eq:FminSyn}
            F_{min} =  1.88 \times 10^{-6}~[~ergs~cm^{-2}~] ~.
        \end{equation}

        \noindent We then apply the Efron-Petrosian statistic to our synthetic data set and find that $\alpha = 0.11 \pm 0.24$ with the detection threshold set at 50\% probability of detection. This is reassuring as it implies that the Efron-Petrosian test statistic remains relatively unbiased even when the detection threshold is not a sharp cutoff. But this is true only for as long as the detection threshold is not significantly underestimated.
        \newpar
         
        To further illustrate this, we consider a lower detection threshold, comparable to the value used in \citetalias{lloyd2019cosmological}. We note that a direct application of the assumed detection threshold of \citetalias{lloyd2019cosmological} to our analysis is not possible since the data set used in \citetalias{lloyd2019cosmological} has been collected from multiple heterogenous sources, as opposed to our synthetic homogenously-detected LGRB sample. However, a detection threshold equivalent to that of \citetalias{lloyd2019cosmological} can be obtained in our analysis by noting that the detection threshold cutoff assumed in the study of \citetalias{lloyd2019cosmological} is above only 3 individual LGRB events. This comprises less than $1\%$ of the entire data set of 376 LGRBs in \citetalias{lloyd2019cosmological}. 
        \newpar
        
        We therefore, choose our detection threshold limit such that only $1\%$ of our synthetic sample falls below the assumed detection threshold hard cutoff, similar to that of \citetalias{lloyd2019cosmological}. This yields a value of $\alpha = 1.55 \pm 0.21$ which depicted illustrated in Figure \ref{fig:synSamFluenceThreshAlpha}. This inferred non-zero correlation at $7\sigma$ significance clearly contradicts the fundamental assumption of our Monte Carlo simulation, and confirms our hypothesis that an underestimation the detection threshold can readily bias the Efron-Petrsoan test statistic.
        \newpar

        Next, we repeat the above analysis for the joint distribution of $z-\liso$ with a detection threshold hard cutoff set at 50\% probability of detection: $1.88 \times 10^{-7}~[~ergs~cm^{-2}~s^{-1}]$. We find $\alpha = -0.04 \pm 0.24$ at this probability of detection. However, when we use a detection threshold comparable to those of \citetalias{yu2015unexpectedly}, \citetalias{pescalli2016rate}, and \citetalias{tsvetkova2017konus}, we find $\alpha = 1.72 \pm 0.16$ at $>10\sigma$ significance, again contradicting the a priori assumption of our Monte Carlos universe of LGRBs.

\section{discussion}
\label{sec:discussion}

    In this work we re-analyzed several previous studies on the evolution of the luminosity/energetics of LGRBs with redshift. To be consistent with the previous studies we used the method of Efron-Petrosian and the $\tau$ statistic to determine the exponent of the power-law relationship, $\alpha$ in $g(z)=(1+z)^{\alpha}$, between the luminosity/energetics and redshifts of LGRBs.
    \newpar

    Contrary to the previous studies, we conclude that the effects of the detection threshold has been likely severely underestimated. We further confirm our conclusion via Monte Carlo simulation, where we assume no correlation between the energetics and the redshift of LGRBs. We then measure these simulated LGRBs via the BATSE detector (for its simplicity). Our finding is that {\bf an underestimation of the effective detection threshold by even less than a factor of two can create artificial correlations between the redshifts and the luminosity/energetics of LGRBs}. The Monte Carlo simulation of \citetalias{pescalli2016rate} also shows that an underestimation of the detector flux limit can lead to apparent artificial correlations between luminosity and redshift. They further show that this effect can also give rise to an apparent overabundance of LGRBs at low redshift.
    \newpar

    The regression slope ($\alpha$, on a log-log plot) of the reported correlations between the redshift and the luminosity/energetics of previous studies also resembles their chosen detection threshold (Figures \ref{fig:Y15zonea}, \ref{fig:P16figsa}, and \ref{fig:L19figsa}). This further corroborates our hypothesis that the observed correlations are an artifact of the individually chosen detection thresholds of the various gamma-ray detectors.
    \newpar

    A more accurate study of the luminosity/energetics-redshift evolution requires a more careful modeling of the detection threshold of gamma-ray detectors, where the detection threshold is not a single cutoff on the distribution of LGRBs but rather a dispersed set of detection probabilities in the entire bivariate distribution. However, such a modeling approach is impossible with the original method of Efron-Petrosian and requires parametric modeling of the luminosity/energetics and redshift distribution as well as the detection threshold.
    \newpar

    \citet{le2020resolving} uses purely parametric methods to determine the GRB formation rate $\rho(z)$. They find that there is no deviation from the SFR at any redshift for the complete unbiased LGRB \emph{Swift}-Perley and \emph{Swift}-Ryan-b samples. They do, however, find an excess at low redshift $(z<1)$ for the \emph{Swift}-Ryan 2012 sample, and conclude that the reason for this excess is either incomplete sample size, that $\rho(z)$ doesn't trace SFR at low redshift, or that it is simply unclear.
    \newpar

    The premise of the previous studies has been to provide a nonparametric investigation of the luminosity/energetics vs redshift. However, upon performing the nonparametric correlation, the majority of these investigations rely on parametric fitting of the luminosity/energetics and redshift distributions, which convolutes the premise. Indeed, \citet{lan2019luminosity} presents a fully parametric study of the redshift/energetics evolution and reports a potential correlation between the two, but nevertheless, their study is founded on the assumption of a simple hard cutoff of the detection threshold of LGRBs. A fully parametric study of the correlation which incorporates a more accurate and detailed description of the detection threshold of gamma-ray detectors remains to be done. 

\bibliographystyle{mn2e}
\bibliography{../../../libtex/all}

\label{lastpage}

\end{document}